\def\hst{{\sl HST}}

\def\wfc3{WFC3}

\def\vlt{{\sl VLT}}

\documentclass[12pt,preprint]{aastex}
\usepackage{appendix}
\usepackage{epsf}
\usepackage{color}
\usepackage{graphicx}
\usepackage{graphics}
\usepackage{epsfig}
\usepackage{longtable}
\usepackage{multirow}
\usepackage{booktabs}
\begin{document}
\title{Near-Infrared Variability Study of the Central 2.3\arcmin$\times$2.3\arcmin\ of the Galactic Centre II. Identification of RR Lyrae Stars in the Milky Way Nuclear Star Cluster}
\author{Hui Dong$^{1}$, Rainer Sch{\"o}del$^1$, Benjamin 
F. Williams$^2$, Francisco Nogueras-Lara$^1$, Eulalia Gallego-Cano$^1$,Teresa Gallego-Calvente$^1$, Q. Daniel Wang$^3$, R. Michael Rich$^4$, 
Mark R. Morris$^4$, Tuan Do$^4$, Andrea Ghez$^4$}%, Zhiyuan Li$^{5,6}$}

%Michael Rich$^4$

\affil{$^1$ Instituto de Astrof\'{i}sica de Andaluc\'{i}a (CSIC), Glorieta de la Astronom\'{a} S/N, 
E-18008 Granada, Spain}\affil{$^2$ Department of Astronomy, Box 351580, University of 
Washington, Seattle, WA 98195, USA}\affil{$^3$ Department of Astronomy, University of Massachusetts,
Amherst, MA, 01003, USA}\affil{$^4$Department of Physics and Astronomy, University of California, Los
Angeles, CA, 90095, USA}%\affil{$^5$ School of Astronomy and Space Science, Nanjing University, 
%Nanjing, 210093, China}\affil{$^6$ Key Laboratory of Modern Astronomy and Astrophysics at 
%Nanjing University, Ministry of Education, Nanjing 210093, China}\affil{E-mail: hdong@iaa.es}

\begin{abstract}
Because of strong and spatially highly variable interstellar 
extinction and extreme source crowding, 
the faint ($K$$\geq$15) stellar population in the Milky Way's 
nuclear cluster is still poorly studied.
%Limited by high and strong spatially variable extinction, 
%the old ($>$10 Gyr) %metal-poor ([Fe/H]$<$-1) 
%stellar population, in the Milky Way 
%nuclear star cluster, has not been well constrained yet. 
RR Lyrae stars provide us with a tool to 
estimate the mass of the oldest, relative dim stellar population. 
Recently, we analyzed \hst/WFC3/IR observations of 
the central 2.3\arcmin$\times$2.3\arcmin\ %($\sim$5$\times$5 pc$^2$) 
of the Milky Way 
and found 21 variable stars with periods between 0.2 and 1d. Here, we present a further comprehensive 
analysis of these stars. The period-luminosity relationship of RR Lyrae %stars 
is used 
to derive their %foreground 
extinctions and distances. Using multiple 
approaches, we classify our sample as four RRc, four RRab and three candidates, %and  
ten binaries. Especially, the four RRabs %stars 
show sawtooth light curves and 
fall exactly onto the Oosterhoff I division in the Bailey diagram. 
Compared to the RRabs reported by 
Minniti et al, 2016, our %four 
new RRabs have higher extinction ($A_K$$>$1.8) 
and should be closer to the Galactic Centre. %Nuclear Bulge. 
The extinction and distance of 
one RRab match those for 
the nuclear star cluster given in previous works. 
We perform simulations and find that after correcting for incompleteness, 
there could be no more than 40 RRabs 
within the nuclear star cluster and in our field-of-view. 
 Through comparing with %the number of RRabs in 
 the known globular clusters of the Milky Way, 
 %we suggest that an old 
% metal-poor (-1.5$<$[Fe/H]$<$-1) stellar population could exist  
 %in the Milky Way nuclear star cluster, but it can contribute at most
%\textbf{4.4$\times10^5~M_{\odot}$, i.e.
% $\sim$17\% of the stellar mass.} 
we estimate that if there exists an old, metal-poor 
(-1.5$<$[Fe/H]$<$-1) stellar population in the Milky Way 
nuclear star cluster on a scale of 5$\times$5pc, then it contributes at most 
4.7$\times10^5~M_{\odot}$, i.e., 
$\sim$18\% of the stellar mass.

%Because of the observational difficulties 
%imposed by strong and spatially highly variable interstellar 
%extinction and by extreme source crowding, 
%the faint ($K$$\geq$15) stellar population in the Milky Way's 
%nuclear cluster is still poorly studied.
%Limited by high and strong spatially variable extinction, 
%the old ($>$10 Gyr) %metal-poor ([Fe/H]$<$-1) 
%stellar population, in the Milky Way 
%nuclear star cluster, has not been well constrained yet. 
%RR Lyrae stars, which are widely found 
%in globular clusters, provide us with a tool to 
%trace and estimate the mass of the dim oldest stellar population. 
%Recently, we analyzed \hst\ WFC3/IR observations of 
%the central 2.3\arcmin$\times$2.3\arcmin\ ($\sim$5$\times$5 pc$^2$) of the Milky Way 
%and found 21 variable stars with 
%periods between 0.2 and 1 d. Here, we present a further comprehensive 
%analysis of the results for these stars. The period-luminosity relationship of RR Lyrae stars was used 
%to derive their foreground extinctions and distances. Using multiple 
%approaches, we classified our sample as four RRc stars, four and three candidate RRab stars and  
%ten eclipsing binaries. Especially, the four RRab 
%stars show sawtooth light curves, have an average period of 0.567 d and 
%fall exactly onto the Oosterhoff I division in the Bailey diagram. 
%Compared to the RRab stars repo
\end{abstract}

\section{Introduction}\label{s:introduction}
Nuclear star clusters (NSCs) have been discovered in 60-70\% of all
types of local galaxies. NSCs typically have sizes similar to globular
clusters, but are 1-2 orders of magnitude brighter. Thus, they are the most massive and
dense stellar systems in the present-day Universe. The masses of
NSCs are correlated with the masses of their host galaxies. NSCs,
normally characterized by complex stellar populations, often show
signs of recurrent and very recent star formation~\citep[see][and
references therein]{bok02,bok04,wal05,wal06,car15,geo16}.

Two mechanisms have been proposed for the formation of NSCs: (1) NSCs
could grow largely {\it in-situ} through gas infall followed by star
formation or by accretion of clusters formed in their close
environment~\citep{aga11,neu11,bek06}. These scenarios are supported by,
among other forms of evidence, the flattening and rotation observed in NSCs of 
edge-on galaxies and by the presence of young ($\leq100\,$Myr) 
stellar populations in some NSCs \citep{wal06}.  In particular,
{\sl in-situ} star formation occurred a few Myr ago in the centre of the
Milky Way and may still be ongoing 
\citep[][and references therein]{gen10,lu13,yus15}. (2) NSCs
may also have acquired a significant fraction of their masses through the
infall and dissolution of globular clusters due to dynamical friction
and tidal forces~\citep[e.g.][]{tre75,lot01,ant12}. Globular clusters
could thus have contributed to the population of the oldest stars. %While
%globular clusters in the Milky Way have generally low metallicities,
%those in the Galactic bulge can reach values of
%[Fe/H]$\approx0$. One may expect that globular clusters that
%created or merged with the NSC were originally located close to the
%Galactic Centre (GC). Therefore, the infall scenario
%cannot be categorically ruled out by recent observations that indicate
%that only about 5\% of the stars in the NSC have [Fe/H]$\leq -0.5$
Around 5\% of the stars in the NSC have been found to have [Fe/H]$\leq -0.5$ 
and may have been 
contributed by old metal-poor globular clusters
\citep[][Rich et al. 2017, in preparation]{do15,fel17,sch16,ryd15}. It is also possible that both 
{\sl in-situ} star formation and globular cluster infall may have contributed to
the growth of NSCs \citep[e.g.][]{har11}. Detailed stellar
population studies of NSCs are required to assess the relative importance of these two
scenarios.

The large distances to extragalactic nuclei limit us to the study of
only their integrated light, which is averaged over scales between a 
parsec and tens of parsecs, is dominated by the brightest stars, and
may even be contaminated by nuclear activity. In contrast, the GC is
located only $\sim$8 kpc from us~\citep{ghe08,gil09,boe16}, or
about a hundred times closer than the second nearest, comparable NSC in the
Andromeda galaxy. Because of this proximity, the GC is the only
nucleus where we can conduct spatially resolved population
studies~\citep[e.g.,][]{sch09,buc09,gen10}.

The Milky Way's NSC (MWNSC hereafter) has a half-light radius of
approximately 4\, pc and a total mass of roughly
2.5$\times10^7~M_{\odot}$~\citep{sch14a,fri16}. The cluster's rotation 
axis is parallel to that of the Galactic Disk~\citep{fel14,fri16}. The cluster is also
known to have a quasi-continuous, complex star formation
history~\citep[see][and references therein]{pfu11}. The most recent
burst of star formation in the MWNSC happened $\sim$3-6\,Myr
ago~\citep[e.g.,][]{pau06,lu13}. %, dominates the overall light
%that we observe, but contributes only
%$\sim$0.2\% of the total mass of the MWNSC
The combination of MWNSC, large Nuclear Stellar Disk and 
Nuclear Molecular Disk is called the Nuclear Bulge, which, in projection, 
appears to be a flat bar with an outer radius of 230 pc and a scale height 
of 45 pc~\citep{lau02}.

While the most recently formed stars are generally believed to have 
formed {\sl in situ} (see, e.g., discussions in~\citealt{gen10}
or~\citealt{lu13}), there is still no compelling evidence for the
presence of a stellar population that may have been contributed by 
globular cluster infall. As a first step, finding such evidence requires
identifying $\sim10 $\,Gyr old stars. % with sub-solar metallicities. 
Subsequently, one could study the distribution and
dynamics of such a population to see whether it can result from 
the infall scenario. Because of the extreme
interstellar extinction and strong source crowding toward the GC, 
current imaging studies with
adaptive optics (AO) assisted 8m-class telescopes can achieve only a 
$\sim50\%$ completeness limit of $K_{s}\approx18.5$
\citep{sch14b,sch17,gal17}. Spectroscopic studies are generally limited to
$K_{s}\leq16$ \citep{pfu11}. The mean mass of the spectroscopically
accessible, $K_{s}=15-16$ Red Clump (RC) giants is $>1\,$M$_{\odot}$
\citep[see Fig.\,16 in][]{sch07},
which means that they may not be old enough to serve as potential
tracers of ancient globular cluster infall.

RR Lyrae (RRL) stars provide us with a %n alternative 
method to study
the old stellar populations ($>$ 10 Gyr old,~\citealt{wal89,lee92,pie16}). 
RRL stars are low-mass core-helium-burning stars with oscillation 
amplitudes in the 0.1$<\delta K_s<$0.5 mag range and 
period range between about 0.2 and 1 d. Although there are metal-rich RRLs 
observed in the
Galactic Disk \citep[see][and references therein]{cha16}, the majority of 
RRLs are found in metal-poor globular clusters~\citep{cat09}. Based on 
the amplitude-period diagram of such RRLs 
(also known as a Bailey diagram), the globular clusters in the Milky Way
can be divided into two types: Oosterhoff type I and II (OoI and OoII,
hereafter). In general, the OoI clusters ([Fe/H]$\sim$[-1.0,-1.5]) seem
to be more metal-rich than the OoII clusters ([Fe/H]$\sim$[-1.5,-2.5])
(Fig. 5 in~\citealt{cat09}). Therefore, by finding RRL stars 
and determining their distribution between these two types, we can  provide 
new constraints on the formation history of the MWNSC.

RRL stars can be classified into three groups: Fundamental-mode
RRLs (Type ab, RRab hereafter), first-overtone RRLs (type c, RRc
hereafter), and the rare double-mode RRLs (type d).  The light curve 
of an RRab star is very unique, with a broad maximum, a sharp minimum,
and a steep ascending branch. In contrast, the light curve of an RRc star 
is typically rather 
symmetric and can be fitted with a single sine function. Also, the periods 
of RRab stars are longer than those of RRc stars; 
they can be distinguished at a period of 0.4\,d~\citep{gra15}. 

The biggest challenge to detecting RRL stars in the MWNSC is their intrinsic 
faintness. Accounting for the appropriate extinction and distance modulus of 
the GC, their observed magnitudes are $K\approx17$ mag and
$H\approx18.5$ mag, below the detection threshold of
the seeing-limited (FWHM$\approx0.7\arcsec$) VISTA Variables in the
Via Lactea Survey (VVV) that consists of multi-epoch near-infrared (near-IR) imaging observations of
 the Galactic Bulge and
southern Disk since 2010~\citep{min10}. Although \citet{min16}
recently reported the
detection of RRL stars in the Galactic Nuclear Bulge, the
locations and relatively low extinctions of these stars suggest that
they are in the foreground of the Galactic Bulge (see discussion in
\S\ref{ss:loca}).

We follow up here on our recent analysis of the Hubble Space
Telescope (\hst ) Wide Field Camera 3 (WFC3) IR observations covering the
central 2.3\arcmin$\times$2.3\arcmin\ ($\sim$5$\times$5 pc$^2$) of the
MWNSC. Thanks to the high angular resolution ($\sim$0.2\arcsec ),
sensitivity, and stability of these observations we have identified 3894 
variable sources \citep{don17} and have further 
derived the periods for 36 of these sources. The periods, as well as observed
magnitudes and colours of several of the sources, indicate that they
are RRL candidates probably located in the Nuclear Bulge or even in the MWNSC.

The rest of this paper is organized as follows. We briefly describe
our observations, data analysis, and steps to calculate the extinction
and distance using the near-IR period-luminosity (PL) relationship of
our RRL candidates in \S\ref{s:obs}. We classify the types of
candidates on the basis of their periods/magnitudes/colors 
and estimate the contamination from eclipsing
binaries in \S\ref{s:result}.  In \S\ref{s:discussion}, we 
discuss the relationship of our identified RRL stars to the Nuclear Bulge and
the MWNSC and explore the limits that the detection of these stars
places on the oldest stellar population in the GC and on the
globular cluster infall scenario. We summarize our results in
\S\ref{s:summary}.

\section{Observations, data reduction and analysis}\label{s:obs}
\subsection{\hst\ dataset and variable source catalog}\label{ss:hst}
Observations, data reduction, and identification of variable sources
have already been described in Paper I.  Here, we only briefly summarize the key
points.

The \hst/WFC3 IR observations in the F127M (1.27 $\mu$m) and F153M
(1.53 $\mu$m) bands that we used 
were from programs GO-11671, GO-12318, GO-12667,
GO-12182, GO-13049, GO-13116, and GO-13403
~\citep{hos15,sto15,mos16}. In terms of their effective wavelengths,
F127M and F153M are the analogs of the the Johnson-Glass J band (1.22
$\mu$m) and H band (1.63 $\mu$m) filters.  The observations in the F153M
band, acquired between 2010 and 2014, include 290 dithered
exposures, ranging from 250 to 350 seconds each. The total length of
the exposures in the first three
years is  less than 6 hours, but the observation blocks in 2014 are long: 10
hours on Feb 28, 10 hours on Mar 10, 15 hours on April 2, and 5.5
hours on April 3. These blocks are critical in our identification of RRL stars, 
because of the significant coverage of their periods. We use the observations
in the F127M band to determine F127M-F153M colours of the stars, which helps to locate them
in the foreground/background or in the MWNSC
proper. The total duration of the F127M observations is 
less than 2 hours. Therefore,
they can only cover a small fraction of the periods of the RRL stars.

We used `DOLPHOT'~\citep{dol00} to detect sources, extract photometry
from individual dithered exposures, and empirically determine the photometric 
variations among exposures.
We further used the least $\chi^2$ method to identify variable
stars and then the Lomb-Scargle periodogram analysis~\citep{lom76,sca82} to
calculate the periods of 36 sources with well-covered light curves
within the individual observation blocks in 2014. We found that 21
sources have periods between 0.2 and 1\,d, and therefore could be RRL
stars. Table~\ref{t:period} lists their
IDs in~\citet{don17}, celestial coordinates,
periods, as well as F127M and F153M magnitudes and  
the corresponding uncertainties. 
%($\sigma_{F127M}$ and $\sigma_{F153M}$). 
%For simplicity, we labeled
%the sources as R1 to R21.  
To simplify the presentation in this paper, we label RRL candidates as R1 to R21.

\begin{figure*}[!thb]
  \centerline{    %   \epsfig{figure=fig/surface_dis_image.ps,width=0.4\textwidth,angle=0}
       \epsfig{figure=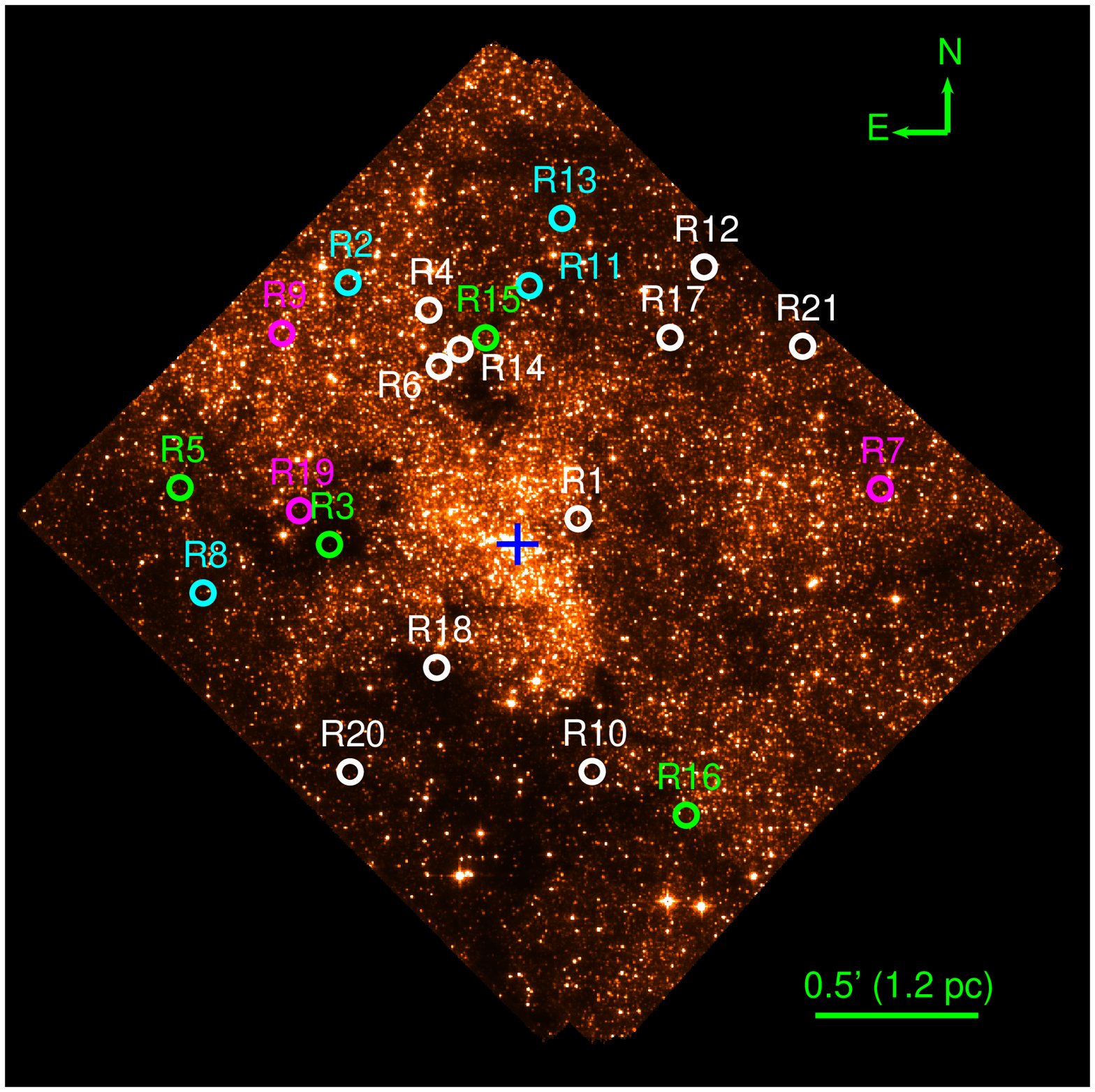,width=1.2\textwidth,angle=0}
       }
 \caption{\hst\ WFC3/IR F153M observations of the MWNSC. The blue plus marks the central 
 massive black hole, Sgr A*. The green circles are the four variables with typical RRab light curves. 
 The cyan circles mark RRc candidates, magenta ones RRab candidates, and white
 ones eclipsing binary candidates.}
\label{f:sgra_f153m}
 \end{figure*}

\begin{figure*}[!thb]
  \centerline{    %   \epsfig{figure=fig/surface_dis_image.ps,width=0.4\textwidth,angle=0}
       \epsfig{figure=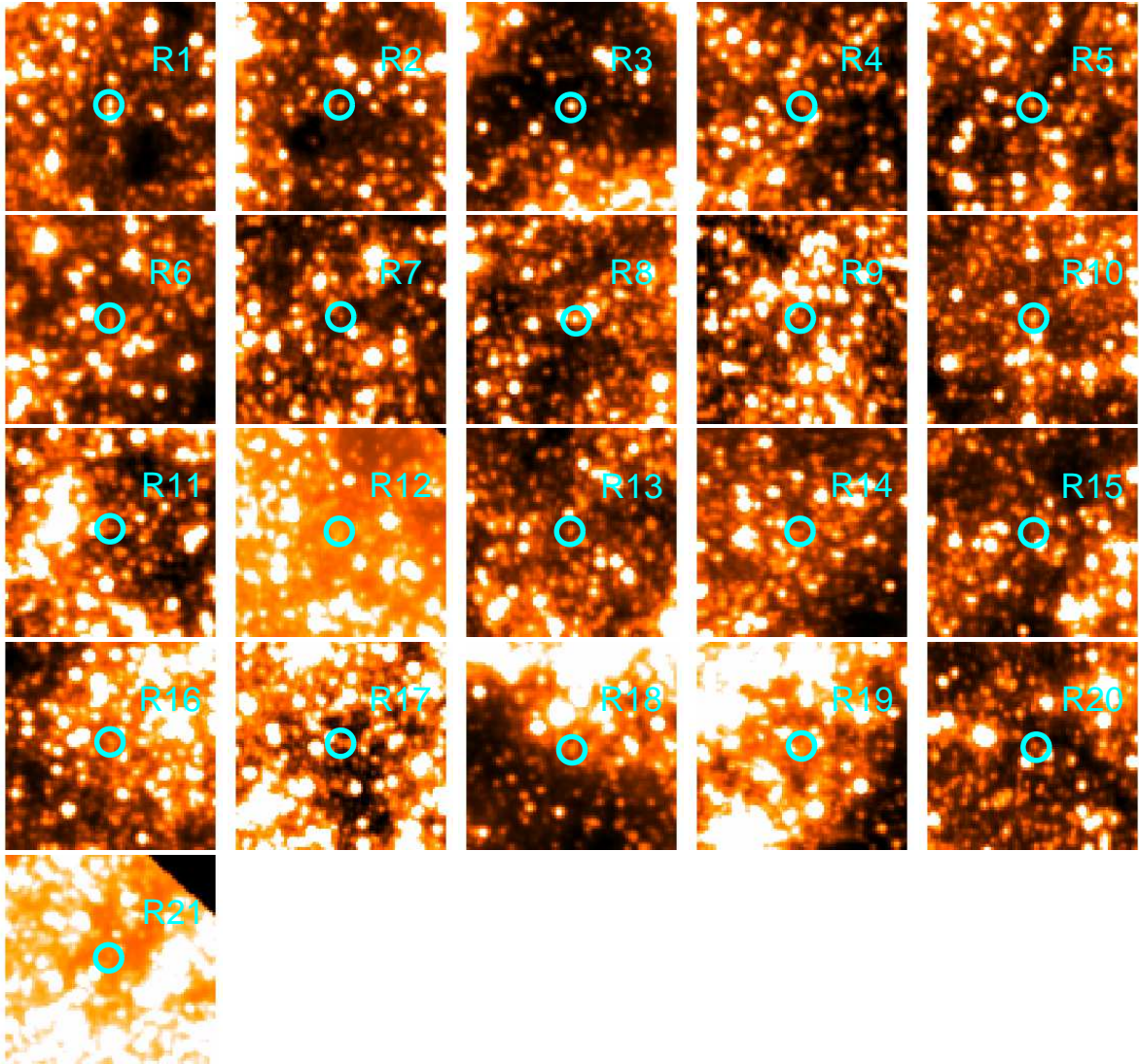,width=1.0\textwidth,angle=0}
       }
 \caption{Detailed images of the surroundings of each star in our sample. The size of each map is 10\arcsec$\times$10\arcsec .%The finding chart at the F153M band of our sample with a size of 10\arcsec$\times$10\arcsec .
 }
 \label{f:finding_chart}
 \end{figure*}

\begin{figure*}[!thb]
  \centerline{    %   \epsfig{figure=fig/surface_dis_image.ps,width=0.4\textwidth,angle=0}
       \epsfig{figure=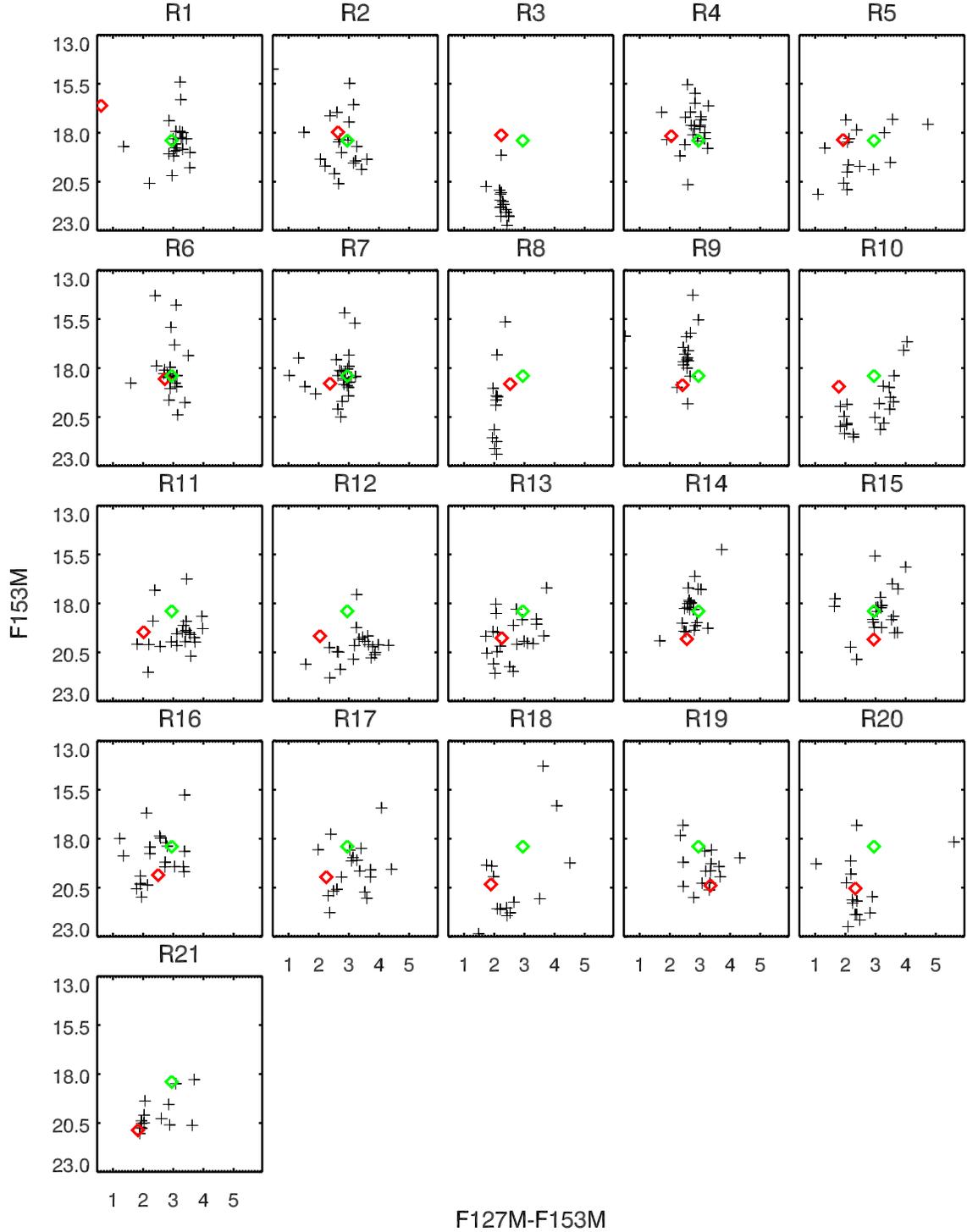,width=1.0\textwidth,angle=0}
       }
        \caption{The colour magnitude diagram (F127M-F153M vs. F153M) of 
 the detected sources (pluses) within 2\arcsec\ of the 21 RRL candidates (red diamonds). 
The green diamonds represent the location of the RC giants in the MWNSC  
 with surface temperature $T_{eff}$=4750$K$, gravity log g=2.5~\citep{puz10}, solar metallicity, 
 $M_K$=-1.54 mag~\citep{gro08}  
 and $A_{K}$=2.5~\citep{sch10}.  %, 
% the IDs of which are given in the title.
}
 \label{f:RRL_can_cmd}
 \end{figure*}

\begin{figure*}[!thb]
  \centerline{    %   \epsfig{figure=fig/surface_dis_image.ps,width=0.4\textwidth,angle=0}
       \epsfig{figure=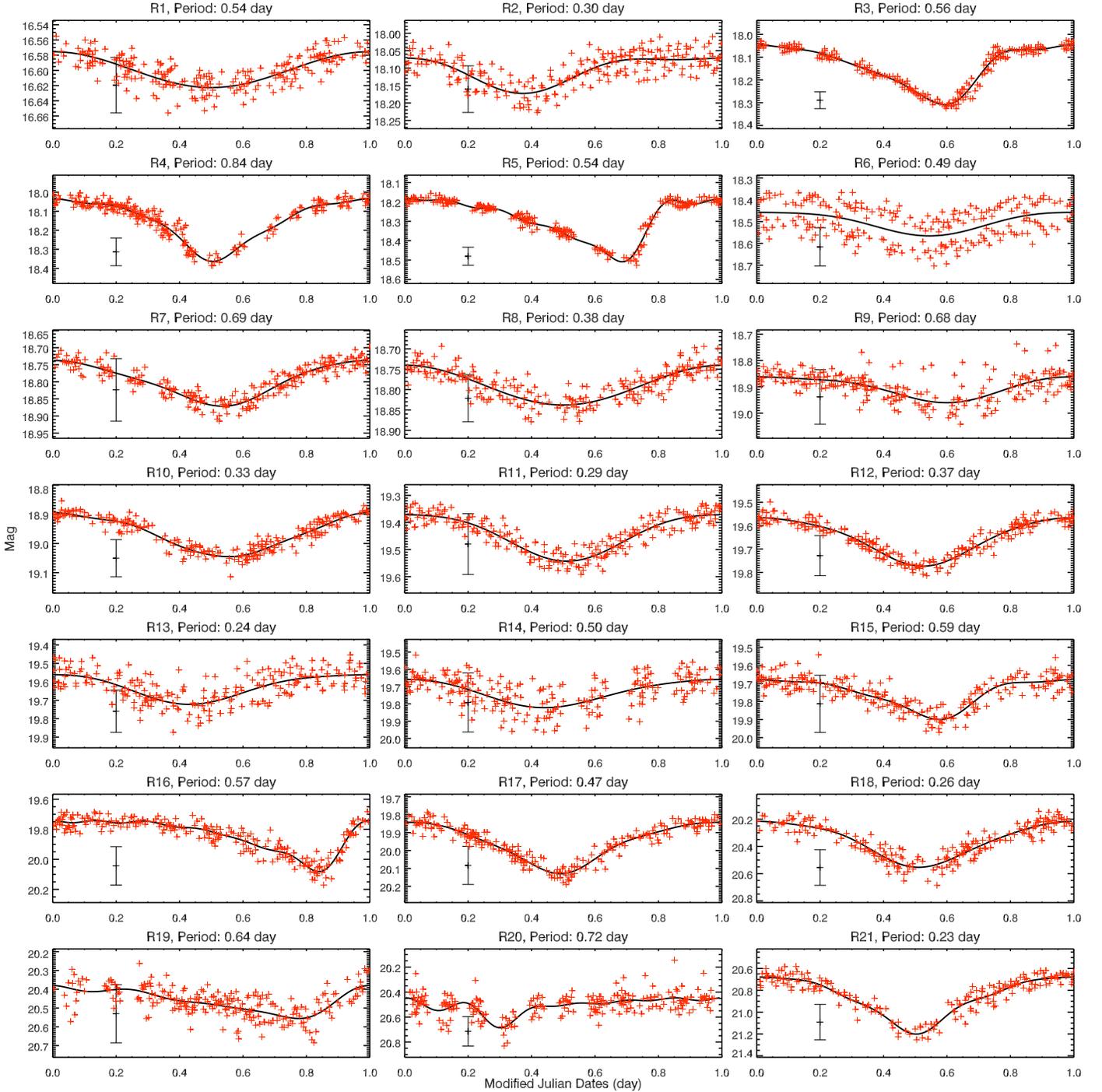,width=1.2\textwidth,angle=0}
       }
 \caption{The folded F153M light curves for 21 sources. 
 In the title of each figure, we give the source IDs and periods. The black pluses
 in the left bottom corner of the individual panels show the average photometric 
 %uncertainties 
 variation among dithered exposures 
 derived from 
 the artificial star tests, plus 0.01 mag systematic uncertainty (see Paper I for more details). The black 
 solid lines are from the DFF fitting.}
\label{f:period_paper2}
 \end{figure*}

Fig~\ref{f:sgra_f153m} shows a mosaic image from the
\hst/WFC3 IR F153M observations overlaid with the positions of the 21
sources and Fig.~\ref{f:finding_chart} gives their finding charts. Fig.~\ref{f:RRL_can_cmd} 
presents the colour magnitude diagram (CMD, F127M-F153M vs. F153M) of 
 detected sources within 2\arcsec\ of each RRL candidate. 
 The folded light curves are shown in Fig.~\ref{f:period_paper2}.
 
\subsection{Fourier fitting of folded light curves}\label{ss:fourier}
We used the direct Fourier fitting (DFF) 
method given in~\citet{kov07} to analyze the folded light curves of these 21 
variable stars. This method fit the data with the following equation: 
\begin{equation}
F153M(\phi)=A_0+\sum_{i=1}^{N}A_{i}\sin(2\pi i\phi+\Phi_i)
\end{equation}
where $\phi$ is the phase of the individual observation, 
and $\Phi$ is the reference phase, and 
$N$ is the maximum order, which is always $\leq$ 6 to prevent over-fitting 
the light curves. 
The definition of the Fourier coefficients in Table \ref{t:rrly} 
are given below
\begin{eqnarray}
A_{i1}=\frac{A_i}{A_1}\\
\Phi_{i1}=\Phi_i-i\Phi_1
\end{eqnarray}
The routine also outputs the $\langle F153M \rangle$ ($A_0$), 
the mean F153M magnitude. From the fitting parameters, we also 
derived the peak-to-peak amplitude, i.e., the difference 
between the maximum and the minimum magnitudes.

\subsection{Extinctions and distances}\label{ss:extin}
We derived foreground extinctions and distances of the RRL
candidates from their apparent F127M/F153M magnitudes,
with the help of the PL relationship of RRL stars.

\begin{figure*}[!thb]
  \centerline{    %   \epsfig{figure=fig/surface_dis_image.ps,width=0.4\textwidth,angle=0}
       \epsfig{figure=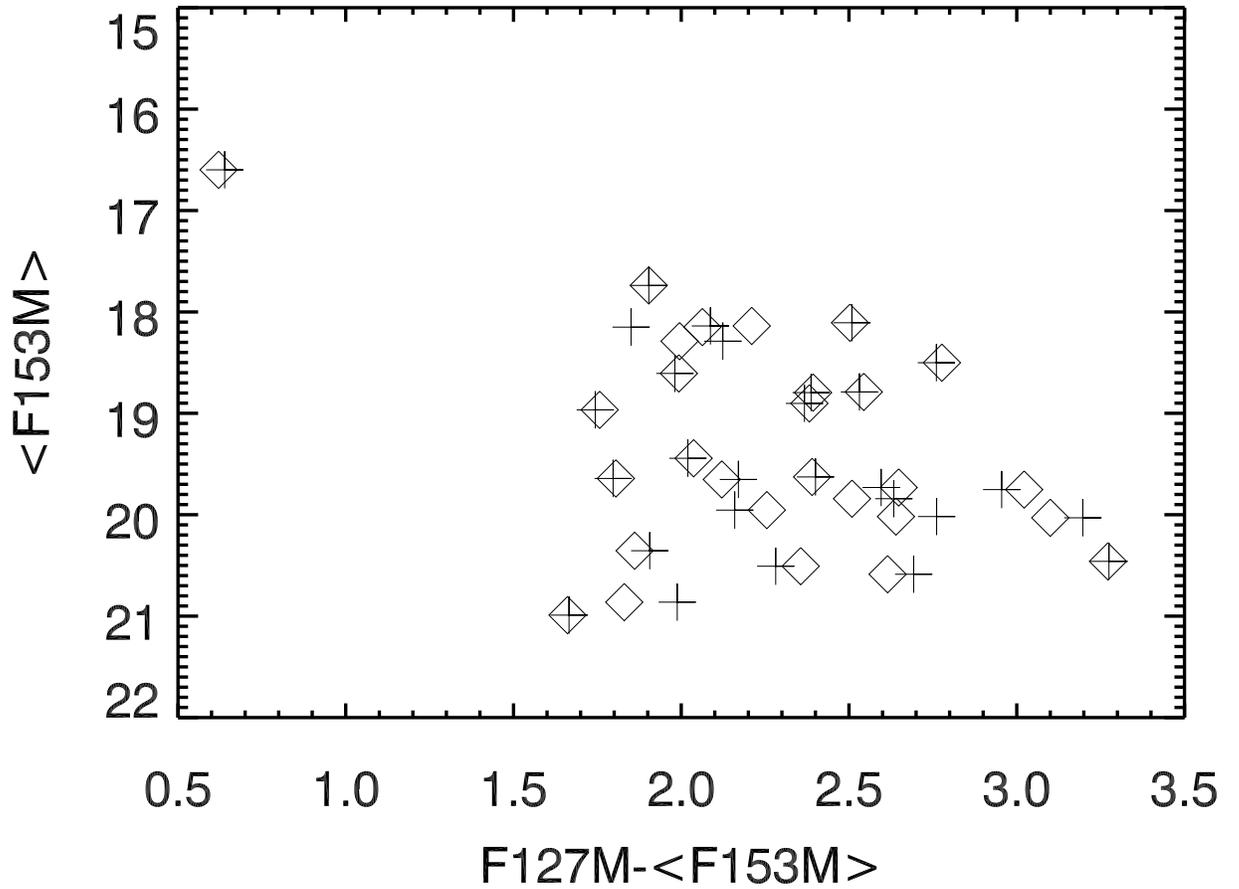,width=1.2\textwidth,angle=0}
       }
 \caption{The (F127M-$\langle F153M\rangle$) vs. $\langle
   F153M\rangle$) CMD for the 21 RRL candidates. The diamond and cross symbols represent the observed 
 F127M magnitude and the mean F127M magnitude, $\langle F127M\rangle$, after the 
 phase correction with the light curves from the F153M band (see more details in \S\ref{ss:extin}).}
\label{f:rrly_mag_col}
 \end{figure*}

Since the observations in the F127M and F153M bands probably cover
different phases of the light curves, the observed F127M-F153M colours
are biased. For each star, we calculated first its average
F127M magnitude, $\langle F127M \rangle$ and then its colour
$\langle F127M \rangle$-$\langle F153M \rangle$.  %While
%$\langle F153M \rangle$ was obtained from the DFF fit (see
%\S\ref{ss:fourier}),
We estimated $\langle F127M \rangle$ using 
the F127M magnitude from the observations of Program GO-11671, and, if
unavailable, Program GO-12182.  The estimate assumed that the 
light curves in the F127M
band are similar to those in the F153M band, but may have 
different amplitudes, used Equations (B3) and (B4) 
in~\citet{fea08}\footnote{\citet{fea08} use $J$ and $H$ in their Equations (B3) and
  (B4). Considering the closeness of the effective wavelengths between
  $J$/$H$ and F127M/F153M (\S\ref{ss:hst}), we assume
  $\Delta$F127M=$\Delta J$ and $\Delta$F153M=$\Delta H$.}: 
\begin{equation}
\Delta F127M = -0.015 + 2.18447\times(\Delta F153M -0.111) 
\end{equation}
This procedure introduces a $\sim$0.1 mag systematic uncertainty into
the $F127M$ magnitude according to~\citet{fea08} and~\citet{gra15}.
From the artificial $F127M$ light curves and the phases of the F127M
observations, we then calculated the difference between the observed
$F127M$ magnitude and $\langle F127M \rangle$, which could be up to
$0.27$\,mag.  The diamonds and crosses in Fig.~\ref{f:rrly_mag_col}
represent the colors, $F127M-\langle F153M \rangle$ (diamonds) and
$\langle F127M \rangle$-$\langle F153M \rangle$ (pluses).  Most of the
candidates, have $\langle F127M\rangle$-$\langle F153M\rangle >1.5$,
except for R1 ($\langle F127M
\rangle$-$\langle
F153M \rangle$=0.58), which we therefore conclude is a foreground object.

Subsequently, we derived the absolute magnitudes ($M_{F127M}$ and
$M_{F153M}$) of our RRL candidates.  Unlike in the ultraviolet and
optical bands, the PL relation for RRL stars in the near-IR is
tight~\citep{lon90} and is not sensitive to metallicity.~\citet{cat04}
give the following dependence:
\begin{eqnarray}
M_J=-0.141-1.773\times log P + 0.190\times log Z \\
M_H=-0.551-2.313\times log P + 0.178\times log Z 
\end{eqnarray}
where $P$ and $Z$ are the period and metallicity, while  
$M_J$ and $M_H$ are the absolute magnitudes in the 
Johnson-Cousins-Glass system.~\citet{cat04} give the following 
conversion between $Z$ and [Fe/H]: log Z = [Fe/H]-1.765, with the 
assumption that the solar metallicity is 0.01716. 
In Appendix A, we derive conversions from $M_J$ and
$M_H$ to $F127M$ and $F153M$ for RRL stars, with uncertainties
of only $0.003$\,mag and 0.006\,mag, respectively.  In principle, the
metallicity of RRL stars can be estimated from their periods and
phases (such as $\Phi_{31}$; ~\citealt{jur96,smo05}).  Unfortunately,
the empirical relationship for such estimation is available only 
in the $V$ and $I$
bands. Therefore, we adopted values %of the
%metallicity of RRL, 
$\langle[Fe/H]\rangle$=-1.02 with a 
dispersion of 0.25 dex from~\citet{pie12}, 
who studied $\sim$17,000 RRL stars in
the Galactic Bulge from the OGLE survey. 
A  similar value, 
$\langle[Fe/H]\rangle$=-1.0
with a dispersion of 0.16 dex, is reported by~\citet{wal91} 
for RRLs in Baade's window ($l$=1.0317,
$b$=-03.9097), not far from the GC, but with substantially lower 
foreground extinction.
%$\langle[Fe/H]\rangle$=-1.02 with a dispersion of 0.25
%dex has been 
%By adopting the value of~\citet{pie12}, we derived $M_{F127M}$ and
%$M_{F153M}$, given in Table~\ref{t:rrly_pro}. 
The derived $M_{F127M}$ and
$M_{F153M}$ are given in Table~\ref{t:rrly_pro}. 
The uncertainties of
$M_{F127M}$ and $M_{F153M}$, which are also included in Table~\ref{t:rrly_pro}, 
account for both
the uncertainty of the metallicity (0.25 dex) and that of the
conversions from $M_J$ and $M_H$ to $F127M$ and $F153M$ given above.

Next, we calculated the interstellar extinction $A_K$ from the
observed magnitudes and the absolute magnitudes:
\begin{equation}
A_K=\frac{(\langle m_{F127m}\rangle-\langle m_{F153m}\rangle)-(M_{F127M}-M_{F153M})}{\frac{A_{F127M}}{A_K}-\frac{A_{F153M}}{A_K}}
\end{equation}
where $\frac{A_{F127M}}{A_K}$ and $\frac{A_{F153M}}{A_K}$ are the relative
extinctions, determined by the extinction law, $A_\lambda\propto\lambda^{\alpha}$~\citep{dra89}. 
Appendix B
gives the method to translate different $\alpha$ to
$\frac{A_{F127M}}{A_K}$ and $\frac{A_{F153M}}{A_K}$ for the RRL
stars. This step could introduce the largest systematic uncertainty
into our analysis. For example, \citet{dek15} and~\citet{mat16} show 
that the different values of $\alpha$ given in the literature can
result in systematic uncertainties of 0.2-0.4 mag in the calculated
distance modulus of Classical Cepheids in the inner part of the
Galaxy. The $\alpha$ towards the GC is known to be different from that
of the solar neighborhood (-1.75,~\citealt{dra89}), but its exact
value is still uncertain. For example, while~\citet{nis06} find
$\alpha = -1.99\pm$0.02,~\citet{gos09} claim that $\alpha$ varies
from one line-of-sight to another, with a mean value and standard
deviation of $-2.64\pm0.52$. 
The two most relevant works on $\alpha$ related to the MWNSC
are~\citet{fri11} and~\citet{sch10}, because they analyze specifically
the extinction curves toward the central parsec of the Milky
Way.~\citet{fri11} use hydrogen emission lines to derive
$\alpha$=-2.11$\pm$0.06 and $A_{2.166}$=-2.62$\pm$0.11 in the inner
14\arcsec$\times$20\arcsec .~\citet{sch10} use the RC 
stars detected within the central 40\arcsec$\times$40\arcsec\ to obtain
$\alpha$=-2.21$\pm$0.24 and $A_K$=2.54$\pm$0.12.  These results agree
well within the uncertainties. For our analysis here, we use
three different values of $\alpha=-2.0,-2.1, -2.2$ to obtain $A_{Ks}$ 
values of the RRL candidates (Table~\ref{t:rrly_pro}). 
The less negative $\alpha$ is, the larger is $A_{Ks}$. Nevertheless,
$A_{Ks}$ values obtained from assuming $\alpha=-2.1$ overlap with those obtained from the 
other two values within their uncertainties.

Finally, we estimate the distance moduli (DM) of the RRL stars from 
the following equations:
\begin{eqnarray}
DM=\langle F153M\rangle-M_{F153M}-A_{F153M}\\ 
     =\langle F153M\rangle-M_{F153M}-\frac{A_{F153M}}{A_K}A_K
\end{eqnarray}
From the DM, we then infer the distances, as well as their uncertainties 
(Table~\ref{t:rrly_pro}).

\section{Classification}\label{s:result}
Based on these properties of the variables, we make 
classifications of them, in particular, separating between 
W UMa and RRL candidates. W~UMa stars are eclipsing 
low-mass overcontact binaries and have periods
between 0.2 to 1\,d, in the same range as RRL stars. They are
intrinsically less luminous than RRL stars. Like RRc stars, their light curves 
are sine-shaped. 
Meanwhile, the light curves of RRab stars could be more sinusoidal in the near-IR 
and have a significantly lower amplitude than in the
visible regime~\citep{cat13}. Therefore, the shape of the observed
light curves by themselves may in some cases not be sufficient to
distinguish between RRab stars with sinusoidal light curves 
(RRab?, hereafter) and W~UMa binaries. In the following,
we will therefore use three steps to distinguish between RRab, RRc and W~UMa
binaries.

\subsection{Periods and light curves} \label{ss:re_pe}
Of our 21 sources, 8 have periods shorter than 
0.4 d and are thus possibly RRc candidates. 
The remaining 13 could be of type RRab.
Especially, in the latter group, R3, R5, R15 and R16 show very typical
asymmetric RRab light curves (see Fig.~\ref{f:period_paper2}).
Instead, the light curve of R4 increases too slowly from the minimum
to the maximum, which is somewhat unusual for a RRab star. 
The folded light curves of R19 and R20 are too noisy for a
reliable classification. 
The other light curves are symmetric (sinusoidal), which makes  
it difficult to decide whether they are RRL stars or eclipsing binaries.

\subsection{Extinction}\label{ss:re_ex}

Comparing the extinctions derived in \S\ref{ss:extin} with an
extinction map derived from the RC stars provides us with another tool
to distinguish between W~UMa and RRL stars. The intrinsic luminosity --
and thus also the distance -- of potential W~UMa stars would have been
overestimated by our use of the PL relationship of RL stars in
\S\ref{ss:extin}. Nogueras-Lara et al. (in preparation) use the RC
stars detected in $JHK$ \vlt/HAWK-I observations of a field of about
$7.5\arcmin\times3\arcmin$ centered on Sgr\,A* to construct an extinction map with
an angular resolution of $\sim$2\arcsec , assuming $\alpha$=-2.2. 
If our sample stars are foreground (background) RRL stars in the
Galactic Bulge with distances smaller (larger) than the GC distance of
7.86$\pm$0.18 kpc~\citep[][see also~\citealt{gil16}]{boe16}, the extinctions derived in
\S\ref{ss:extin} should be significantly smaller (larger) than the
values of this extinction map at their corresponding locations. These
extinction values are also listed in Table~\ref{t:rrly_pro}.

\begin{figure*}[!thb]
  \centerline{    %   \epsfig{figure=fig/surface_dis_image.ps,width=0.4\textwidth,angle=0}
       \epsfig{figure=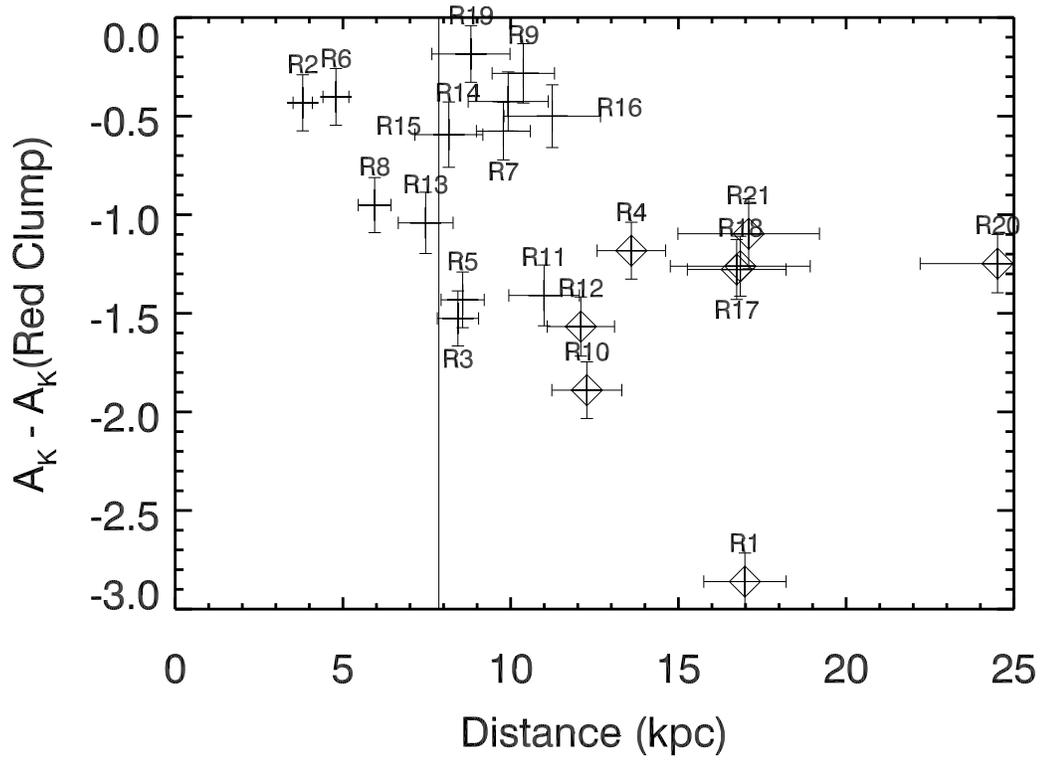,width=1\textwidth,angle=0}
       }
 \caption{The distances derived in \S\ref{ss:extin} vs the differences between the $A_K$ derived 
in \S\ref{ss:extin} and those from the extinction map given in Nogueras-Lara et al. (in preparation). 
The diamonds represent potential eclipsing binaries due to their large distances. 
The vertical line marks the location of the MWNSC~\citep{boe16}.}
\label{f:dis_ext_com}
 \end{figure*}

A plot of the differences between the extinction values estimated in
\S\ref{ss:extin} and those from the extinction map of Nogueras-Lara et al.
(in preparation) plotted over the calculated distances in
\S\ref{ss:extin} is shown in Fig~\ref{f:dis_ext_com}. The calculated
extinction values of all our sample are smaller than those from the
extinction map. This is because 1) stars in front of the GC with
higher extinction than those from
 the extinction map are not expected and 2) the dense molecular clouds in the 
 circumnuclear ring~\citep[][and references therein]{lau13} dim background 
RRL stars and W UMa binaries, which causes them to be undetected or be 
unclassified as variable stars due to larger photometric uncertainties.  

All stars in our field behind the GC have to suffer at least the interstellar
extinction toward the GC. It is unphysical for a star to lie at a
greater distance while at the same time suffer a lower interstellar
extinction, in particular, because we can reasonably assume the
presence of interstellar dust on the far side of the GC, too. We
therefore assume that stars that show this discrepancy are W~UMa
binaries misclassified as RRLs. We therefore classify eight stars 
(R1, R4, R10, R12, R17, R18, R20, R21) 
with distances derived in \S\ref{ss:extin} with $\alpha$=-2.2 larger 
than 8 kpc by more than three times the corresponding distance uncertainties 
as potential eclipsing binaries. %From the CMD within 2\arcsec\ 
%of these sources given 
Fig.~\ref{f:RRL_can_cmd} shows that their  
$F127M$-$F153M$ colours are similar or bluer than the majority of nearby sources, 
which should be within the Galactic Nuclear Bulge, and are  
inconsistent with the possibility that they are background RRLs. 

\subsection{Bailey diagram}\label{ss:re_ba}
\begin{figure*}[!thb]
  \centerline{    %   \epsfig{figure=fig/surface_dis_image.ps,width=0.4\textwidth,angle=0}
       \epsfig{figure=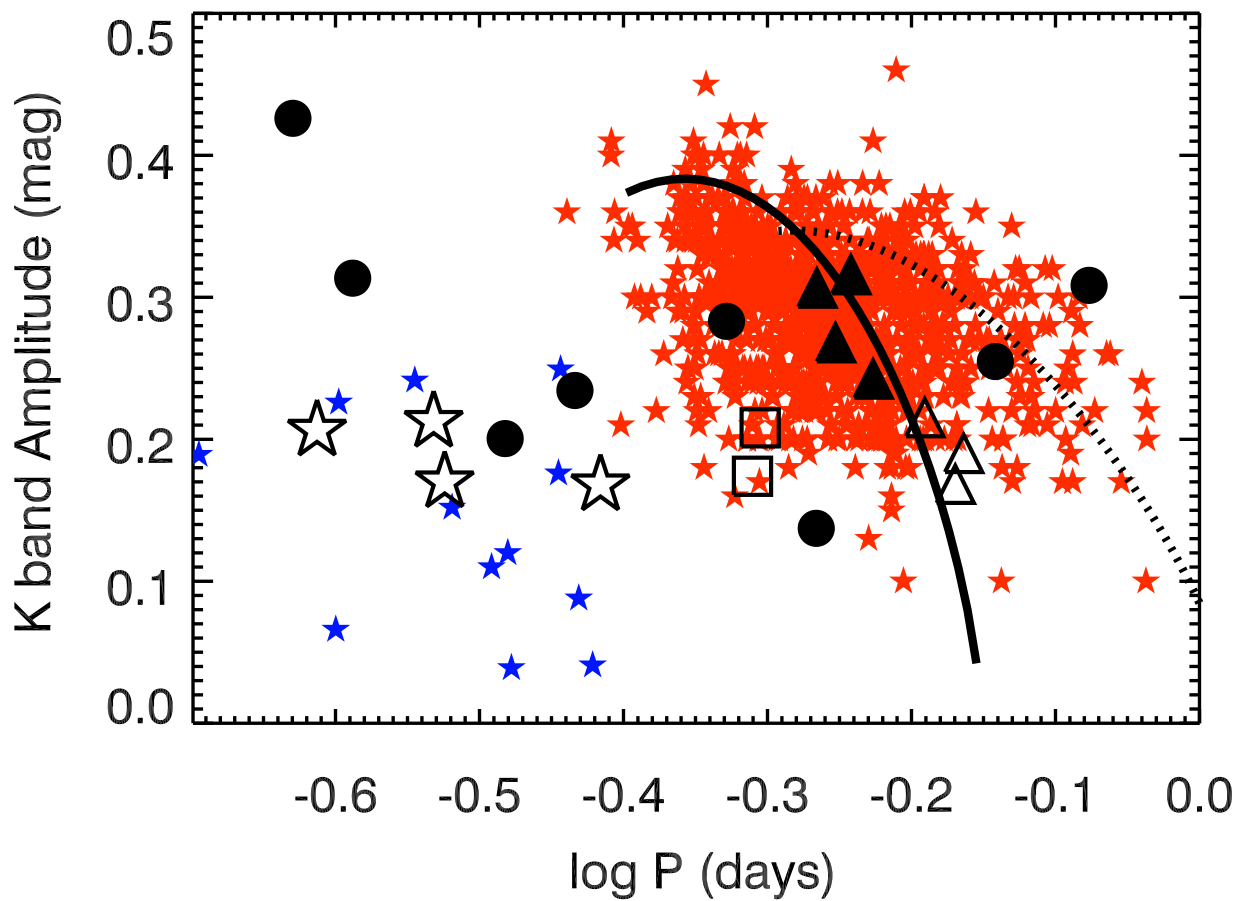,width=1.2\textwidth,angle=0}
       }
        \caption{Bailey Diagram: K-band amplitude plotted against the
         logarithm of the period (d) of RRL stars. The small red and
         blue stars represent the RRab and RRc stars detected in the
         Galactic Bulge by the VVV survey~\citep{gra15,gra16}. The
         large symbols are our variable stars with periods between 0.2
         and 1 d: Stars for RRc candidates, filled triangles for identified RRab
         stars, open triangles for 
         RRab candidates, filled circles for identified eclipsing binary
         candidates, and two open
         boxes for stars that may be RRab or eclipsing binaries.  %The
         %arrows indicate the location of the stars if the amplitudes
         %were not corrected from F153M to $K_s$
         %(Eqn.~\ref{e:amp_cor}).  
         The black solid and dashed lines are
         the OoI and Oo II lines from~\citet{nav15}. }
\label{f:rrlyrae_p_amp}
 \end{figure*}

%In this subsection, we compare the period and amplitude distribution of RRL stars 
%identified from the near-IR VVV survey~\citep{min10} with our candidates.
Fig.~\ref{f:rrlyrae_p_amp} is the $K_s$
band Bailey diagram for our sources along with the
data for a reference sample of RRab~\citep{gra16} and RRc
stars~\citep{gra15}, derived from the VVV survey. Equations (B4) and (B5) in~\citet{fea08} show how
to translate the F153M amplitudes into the $Ks$-band for 
the RRL stars. 
\begin{equation}\label{e:amp_cor}
\Delta Ks = 0.176 + 0.606796\times(\Delta F153M -0.111) 
\end{equation}
%The amplitudes of eclipsing binaries should not show any wavelength
%dependence, of course. 
We can see that 1) the RRc and RRab stars from
our sample lie perfectly within the range for these RRL stars from the VVV survey. 
Especially, the seven RRab stars and
candidates are located very close to the OoI line.  2) Probable W~UMa
binaries as identified in \S\ref{ss:re_ex} are widely distributed
throughout the Bailey diagram. 3) Finally, two stars, R6 and R14 (open
boxes), are located near the edge of the cloud of RRab
stars. Therefore, they may be RRabs or eclipsing
candidates. %, but cannot be classified unambiguously.

To summarise, our 21 sources include four RRc stars, four RRab stars,
three RRab?s, as well as ten probable eclipsing binaries.

\section{Discussion}\label{s:discussion}
Here, we discuss
the physical relationship between RRL stars in our sample and the 
MWNSC, estimate the total number 
of RRab stars in the MWNSC and constrain the 
fraction of the old stellar population. % in \S\ref{ss:origin}. 

\subsection{Physical locations of the detected RRL stars}\label{ss:loca}
In order to obtain constraints on the old stellar population in the
MWNSC, we need to identify which of our identified RRL stars are part of 
the MWNSC. %Unfortunately, since our line-of-sight passes through the
%Galactic bulge, our sample probably contains RRL stars in the
%foreground and possibly also in the background of the MWNSC.  Therefore, 
We first 
use the extinctions and distances determined in \S\ref{ss:extin} to constrain
the line-of-sight locations of our RRL stars with respect to the MWNSC.

Within our sample, three RRL stars lie close to Sgr A* along the line of sight within less 
than the distance uncertainties derived in \S\ref{ss:extin} and 
have $A_K$$>$2: R13, R15 and R19, 
according to their distances with $\alpha$=-2.2 (see Table~\ref{t:rrly_pro}). 
The $A_K$ values of R13 and R15 are smaller than the corresponding values from 
the extinction map of Nogueras-Lara et al. (in preparation), by at 
least five sigma, while the $A_K$ of R19 is consistent with 
the extinction map within two sigma. 
On the other hand, the $F127M$-$F153M$ colours of these stars are similar to those of 
the majority of nearby sources in Fig.~\ref{f:RRL_can_cmd}. Therefore, as suggested 
by their distance,  
R13 and R19 could be naturally explained by an RRc star and an  
RRab? in the foreground and near background of the Galactic Bulge, respectively. 
For R15, an RRab, 
its $A_K$ is still within the uncertainty range of the mean extinction of the 
central parsec ($\sim$25\arcsec ), 2.54$\pm$0.12 mag~\citep{sch10}. Therefore, 
R15 could lie in the Nuclear Bulge, 
close to Sgr A*, but we cannot say with certainty 
whether it belongs to the MWNSC.

\begin{figure*}[!thb]
  \centerline{    %   \epsfig{figure=fig/surface_dis_image.ps,width=0.4\textwidth,angle=0}
       \epsfig{figure=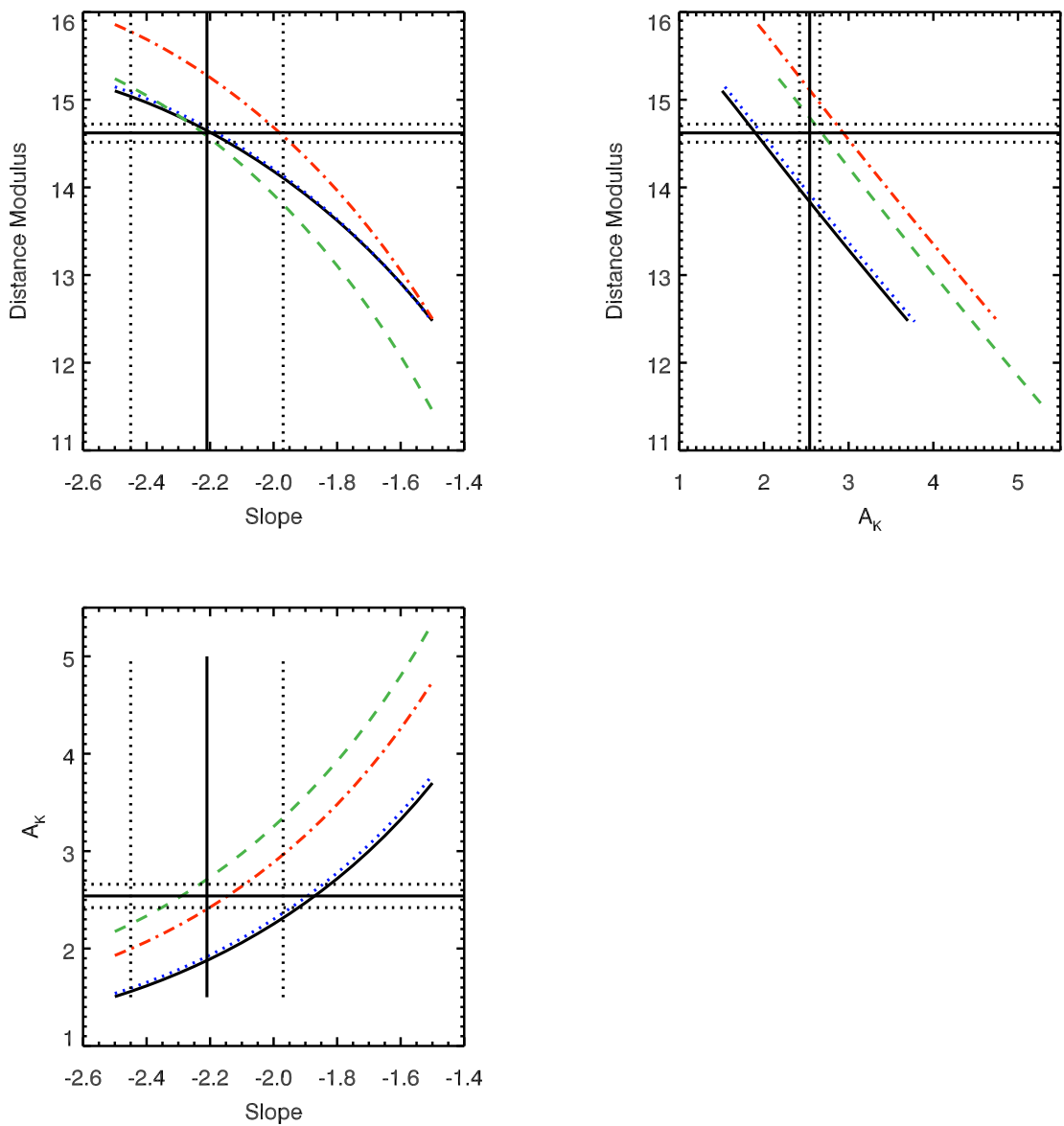,width=1.0\textwidth,angle=0}
       }
 \caption{The correlation between distance modulus, absolute extinction $A_{Ks}$ and the slope 
 of the extinction law ($\alpha$) for four candidate RRab stars (R3: black 
 solid curve, R5: blue dotted curve, R15: green dashed curve, R16: red dot-dashed curve.)
 The horizontal and vertical solid and dashed lines in the three panels are the mean and 1 sigma uncertainty 
 of the distance modulus, $A_{Ks}$ and $\alpha$ of MWNSC given in~\citet{boe16} and~\citet{sch10}. }
\label{f:rrly_slope_ak_dis}
 \end{figure*}

On the other hand, we notice that R3 and R5, two RRabs, appear to 
be located at the GC within their respective distance uncertainties, but with 
$A_K$ smaller than the corresponding values in the extinction map 
of Nogueras-Lara et al. (in preparation) by more than 14 sigma. 
From Fig.~\ref{f:sgra_f153m} and 
Fig.~\ref{f:finding_chart}, we can see that 
R3 and R5 fall into the low stellar number density regions, which suggest the existence 
of dense molecular clouds in front of the MWNSC. Therefore, the extinction of R3 and R5 
indicate that they should be in front of the dense molecular clouds, and then the MWNSC. 
In order to further investigate this dilemma, we give the relationship 
between DM, $A_{Ks}$ and $\alpha$ in Fig.~\ref{f:rrly_slope_ak_dis} for the 
four RRab stars, R3, R5, R15 and R16. 
The DM is anti-correlated with the slope and $A_K$. Therefore, one possible solution 
is that the R3 and R5 are indeed foreground and the extinction 
which they suffer follows a law more similar to that of the solar neighborhood 
rather than a steep extinction law with $\alpha$=-2.2. 
From 
Table~\ref{t:rrly_pro} and Fig.~\ref{f:rrly_slope_ak_dis}, we can see that for $\alpha$ less 
negative than -2.0, 
R3 and R5 could still have extinctions smaller than the values given in~\citet{sch10} and 
be closer to us than the MWNSC. 

The distances of the other six RRL stars indicate that they are at the foreground or background of 
the MWNSC. R2 and R8, having extinctions smaller than those in the extinction map, 
are RRcs in front of the MWNSC. R7, R9, R11 and R16 are behind the MWNSC, but have 
smaller extinction than indicate by the extinction map. In particular, R16 is a typical RRab star. A 
possible explanation is the clumpiness of the molecular gas in the circumnuclear ring. 

%In summary, there are at least three well-defined RRab stars (R3, R5 and R16), 
%which belong probably to the Galactic bulge. We compare
%this number with the value statistically predicted by the VVV
%survey.~\citet{gra16} identified 1019 RRab stars in $\sim$47 square
%degrees in the outer Galactic bulge (-10.0$^o<$l$<$10.7$^o$ and
%-10.3$^o<$b$<$-8.0$^o$). The median and standard deviation of their
%periods are 0.55\,d and 0.087\,d.  Using the location of the RRabs
%in~\citet{gra16}, we produced their surface density as a function of
%the galactocentric radius (`r'), which is then fitted with a S\'ersic
%law, i.e. surface density $\propto$ exp ( k $\times r^{1/n}$ ), where
%k is a constant (the exponential and the de Voucouleurs law correspond
%to the S\'ersic law with n = 1 and 4). Fig.~\ref{f:rrlyrae_surface}
%shows the results for n=1, 2 and 4. We found that for n=4, a de
%Voucouleurs law, the Galactic bulge could contribute $\sim$2.1 RRab
%stars, which agrees with the number 
%found by us, while n=1 or 2, just predict 0.09 or 0.26 RRab
%stars. This also indirectly supports the result of~\citet{dek13} that
%RRL stars in the Galactic bulge seem to follow a classical bulge
%structure ($n=4$): For a pseudo-bulge structure with n $<$ 4, there
%will be too few bulge RRL stars in our field-of-view.

\citet{min16} found 14 RRL stars within the central 36\arcmin\ (86 pc)
from the VVV survey in the $J$, $H$ and $K$ bands. Assuming that
the intrinsic color of RRL stars is ($J$-$K$)$_o$=0.15 mag, they derived
the extinction (E($J$-$K$)). Then they used different versions of
extinction curves and PL relationships for RRL stars to derive their
distances. They claimed that 12 of the RRL stars lie within the Nuclear Bulge. 
However, we notice that the extinction values of
their RRLs range between $A_{K_{s}}=1.06$ and $1.8$, which are 
significantly smaller than what we expect for the Nuclear Bulge. For
example,~\citet{don11} produce an extinction map with 4\arcsec\
resolution from \hst/NICMOS observations covering the central
39\arcmin$\times$15\arcmin . The median and 68th percentile of $A_K$
derived from this map are 2.3 and [2.0, 2.75]. Also, the extinction map
of the Nuclear Bulge presented by \citet{sch14a} suggests values
$A_K>2.0$ for most of the Nuclear Bulge. According to Tables\,1 and 2
of ~\citet{min16}, only five of their RRLs fall in projection onto the
region of the Nuclear Stellar Disk and only two of them (IDs, 37068
and 33007) are within 500 pc of Sgr A* along the line-of-sight. The
$A_K$ values of these two stars are 1.23 and 1.44 mag. Therefore, it is more
likely that these RRL stars are part of the foreground Galactic
Bulge, instead of the MWNSC.

\subsection{RRL  population  in the MWNSC?}\label{ss:total}
Although we have identified only one well-defined RRab, R15, 
that could plausibly belong to the MWNSC, we expect that the MWNSC  
potentially contains a much larger number of such stars that 
are not detected in our study. In
Paper\,I, we divided our field-of-view into ten regions, according to
the local surface brightness. R15 falls into the \#5 region (the
larger the number, the higher the surface brightness) with a 50\%
completeness limit of $F153M=19.2$ mag. RRL stars with similar
magnitudes would not be detected in regions \#7-10, with 50\% completeness
limits $<$ 18 mag. %, they cannot be detected by our method. This does
%not mean, however, that the number of RRL stars in the regions with
%lower surface brightness is complete. On the one hand, 
In addition to the source confusion, the strong extinction is another 
limiting factor for the detection. On average, the
lower surface brightness represents regions with 
greater foreground extinction. 
%which dims RRL stars, too, and the variability of the extinction may
%mean that 
%
%There are patches with too high extinction to detect any 
%RRLs located in the MWNSC. 
This will make it extreme difficult to detect RRLs that are located 
in the MWNSC and are behind the clouds that are responsible 
for the high extinction.” 
Of course, our estimate of the RRL population in the MWNSC 
also needs to consider the stellar spatial distribution of the cluster, 
relative to the foreground and background across the field.

%Finally,
%the low surface brightness regions lie relatively far away from
%Sgr A*, where the number density ratio between stars belonging to the
%MWNSC to those that do not decreases rapidly.
% On the other hand, we have to apply the completeness correction
%factor to the detections in the surface brightness regions. 

\begin{figure*}[!thb]
  \centerline{    %   \epsfig{figure=fig/surface_dis_image.ps,width=0.4\textwidth,angle=0}
       \epsfig{figure=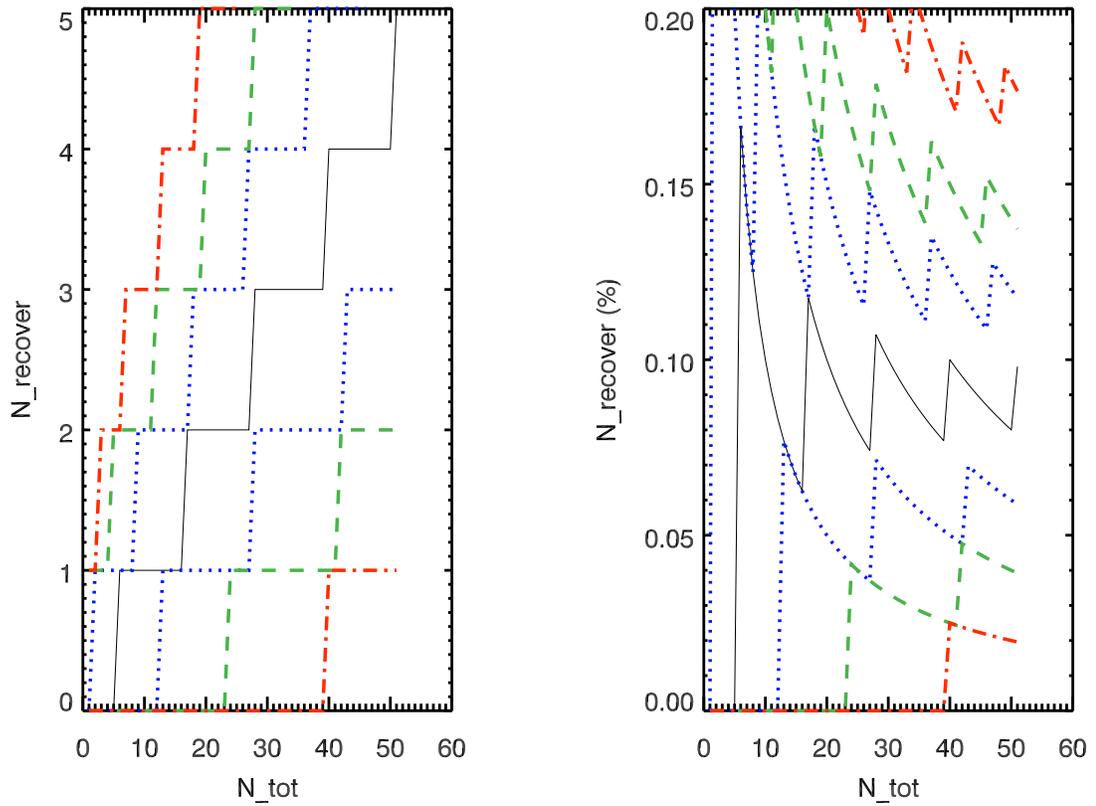,width=1.0\textwidth,angle=0}
       }
 \caption{The recovered number (left) and fraction (right) of simulated RRab stars 
 in the MWNSC. The black solid, blue dotted, green dashed and red 
 dot-dashed lines represent the median, 68\%, 90\% and 99\% percentile. }
\label{f:rrlyrae_simu_recover}
 \end{figure*}

We performed simulations to estimate the detection fraction of RRab
stars in the MWNSC.  We assumed that the distribution of RRL stars
follows the surface brightness distribution as described by Eqn. 1
of~\citet{fri16} with their Model No 4  parameters in their Table 1\footnote{This equation 
includes also the contribution from the Galactic Nuclear Disk, but which is small 
in our field-of-view.}. Then,
we added the distance modulus to the absolute magnitude of R15 and
applied the appropriate extinction. %, derived from its period and
%the PL relationship given in~\citet{cat04}. 
We adopted the R15's light curve as a template. 
In order to simulate the observations, we randomly chose starting
phases of the light curve and assigned the photometric errors
by using the information giving in \S\ref{ss:hst}. The completeness 
as a function of input magnitude from the artificial star tests applied to the 
ten regions (Fig. 5 of Paper I) 
was used to determine whether the RRL stars can be 
detected or not. We then applied the
least-chi-square method described in Paper\,I to check whether each 
detected source may be identified as a variable. We
performed 10,000 such simulations for each of total `N\_tot' RRab stars. %For each N\_tot, we
%ran the simulation 10,000 times. 
Fig.~\ref{f:rrlyrae_simu_recover} shows the median 
fraction of the recovered variable stars, as well as its 65\%, 90\%, 99\%
percentiles. %, which are shown in 
From this plot, we conclude that the median detection fraction is
$\sim$0.1 and exclude N\_tot$>$40 with 90\% confidence.

%we can see that the median detection fraction is
%$\sim$10\% and we can exclude N\_tot$>$40 with 90\% confidence.

\subsection{Old stellar population in the MWNSC?}\label{ss:origin}
 Finally, we use the number of RRab stars to estimate the potential
contribution of an old, metal poor population to the MWNSC, 
which could have been brought into the GC by the infall of globular 
clusters.~\citet{mat09} show
that the numbers of RRab stars in dSph satellite galaxies of the Milky
Way and M31 are strongly correlated with the galaxies' absolute magnitudes. For
a constant mass-to-light ratio, we also expect a tight correlation between
the number of RRab stars and the total stellar mass of dSph galaxies.
 
\begin{figure*}[!thb]
  \centerline{    %   \epsfig{figure=fig/surface_dis_image.ps,width=0.4\textwidth,angle=0}
       \epsfig{figure=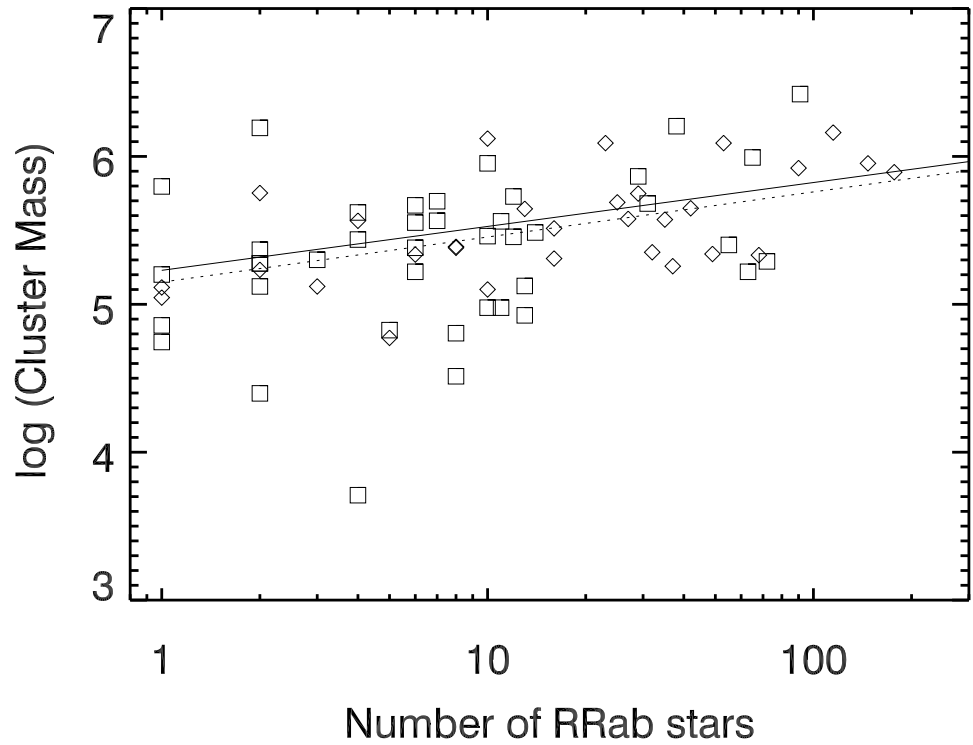,width=1.0\textwidth,angle=0}
       }
 \caption{The relationship between the number of RRab stars and the logarithm of 
   cluster mass in units of solar mass for the OoI (diamonds) and OoII (squares) clusters. 
   The solid (OoI) and dashed (OoII) lines represents the Eqns. 11 and 12, respectively.}
\label{f:rrlyrae_globular}
 \end{figure*}

We tested this relation on Milky Way globular clusters. We used the
variable star catalog for globular clusters given
by~\citet{cle01}\footnote{online catalogs:
  http://www.astro.utoronto.ca/$\sim$cclement/read.html. The catalogs
  have been continuously updated by the 
  authors. We used the data updated on March 22, 2017.}. We only selected 
sources with variable types equal to `R0', `RR0', `RR0?' (all are RRab stars,
according to the draft classification for the 2006 version of 
General Catalog of Variable Stars). The total masses and metallicities  
of globular clusters were taken from~\citet{gne97} and~\citet{har96}, respectively. 
We divide the globular clusters into OoI and OoII clusters according to 
their metallicities. For these two groups, we found 30 and 41 globular 
clusters in~\citet{cle01} with measured total stellar masses and detected 
RRab star(s)\footnote{Four and six clusters in these two groups do not 
have detected RRab star for two reasons: 1) small total masses: 
NGC 6325 (9.6$\times10^4~M_{\odot}$), 
Palomar 4 (5.4$\times10^4~M_{\odot}$) and 
NGC 6287 (1.0$\times10^5~M_{\odot}$)
 and 2) no 
available high-quality time-domain studies: NGC 6218, NGC 6517, 
NGC 4372, NGC 5694, NGC 6144, NGC 6254 and NGC 6752.}. 
 Fig.~\ref{f:rrlyrae_globular} shows the correlation between the 
 total cluster masses and the numbers of RRab stars for these 
 two groups of clusters, although the scatters are large. 
Based on these star clusters, we obtained the following relationship: 
\begin{eqnarray}\label{e:cluster}
log (cluster~mass (M_{\odot})) = 5.13\pm0.12 + (0.33\pm0.09)\times log (number~of~RRab) ; for OoI \\
log (cluster~mass (M_{\odot})) = 5.05\pm0.15 + (0.34\pm0.14)\times log (number~of~RRab) ; for OoII
\end{eqnarray}
with a standard deviation of 0.29 dex and 0.48 dex, respectively. 

By using the relations above, we could estimate the contribution of 
infall globular clusters in the MWNSC. The Bailey
diagram in Fig.~\ref{f:rrlyrae_p_amp} shows that the four likely
RRab stars and three RRab candidates lie close to the OoI line, like most
of the RRab stars in the Galactic Bulge~\citep{gra16}. This means that
our newly identified RRL stars probably have metallicities -1.5$<$[Fe/H]$<$-1.  
Therefore, using the relationship above for the OoI clusters, we could estimate that 
10 and 40 RRab stars -- our best estimated number 
and the 90\% upper limit for the MWNSC, respectively -- correspond to 
globular clusters with masses of 2.9$\times10^5$ [2.1$\times10^5$, 4.1$\times10^5$] M$_{\odot}$ and 
4.7$\times10^5$ [3.1$\times10^5$, 7.2$\times10^5$] M$_{\odot}$ 
(68\% uncertainty intervals). We can compare this value with the stellar mass
of the MWNSC in our field-of-view, $\sim$2.6$\times 10^6~M_{\odot}$, as estimated from Table 
8 of~\citet{fri16}.
%in our field-of-view is 
%According to our estimations above, we conclude that there are not
%more than 40 RRL stars in the MWNSC with a 90\% confidence limit. This
%suggests 
We conclude that the old metal-poor population contributed by the 
OoI cluster occupies at most %at most
18\% of the total mass of the MWNSC in this region, because 
both infall globular cluster and ancient {\it in situ} star formation could 
provide these old metal-poor stars. This is
consistent with the low fraction ($\sim$5\%) of metal-poor stars 
(-1.5$<$[Fe/H]$<$-0.5) 
in the MWNSC found by~\citet{do15} and~\citet{fel17}.  

On the other hand, 
we do not detect any RRab stars, which fall into the OoII region in 
the Bailey diagram. 
These RRab stars have slightly 
longer periods (the average period is $\sim$0.65 d,~\citealt{cat09}) 
than the ones in OoI clusters (the average period 
is $\sim$0.65 d). Therefore, if they existed in the MWNSC and 
were detected in our dataset, their periods could be accurately 
determined. The existence of OoII clusters 
in the MWNSC is also inconsistent with the minimum stellar abundance 
reported by~\citet{do15} and~\citet{fel17}: -1.27 dex and -1.25 dex.  
According to Fig.~\ref{f:rrlyrae_simu_recover} 
and the relationship above for the OoII cluster,  the 90\% upper limit  for the 
total mass of the OoII clusters is 3.3$\times10^5$ M$_{\odot}$ , 
i.e. 13\% of the total mass 
of the MWNSC. %In total, the OoI and OoII clusters must be less than 

\section{Summary}\label{s:summary}
In this paper, we have performed a study of 21 variable stars with
periods between 0.2 and 1 d identified in \hst\ WFC3/IR F153M
observations of the Milky Way nuclear star cluster~\citep{don17}. We have analyzed their light curves and studied
their extinctions and distances. Based on these analyses, 
we have typed the variable stars, estimated their physical 
locations, and discussed the implications of the results, 
focusing on the RRL stars in the MWNSC. Our findings are the following:

\begin{itemize}
\item The 21 sources are classified as: four RRc stars, four RRab stars, three RRab candidates 
and ten eclipsing binaries. In particular, our RRab stars are $\sim$ 2 mag dimmer 
and should be closer to the Galactic Nuclear Bulge 
than those identified in~\citet{min16}.

\item  Using the well-defined period-luminosity function from the literature, 
we calculate the line-of-sight distances of our 11 RRL stars to be 
$\sim$4 kpc to 11 kpc away from us. 

%To determine their line-of-sight distance, we have calculated
 % the absolute extinction and distances of 16 RRL stars and candidates
%  with the help of the period-luminosity function for RRL stars. 

\item All four well defined RRab stars and three RRab candidates 
fall onto the Oosterhoff I division,
  which suggests that the old stellar population near the GC is
  relatively metal-rich with -1.5$<$[Fe/H]$<$-1  
  and could have the same origin as the old
  stellar population in the Galactic Bulge.

\item We have found that only one out of 
our four well-defined RRab stars may actually 
belong to the MWNSC or lie in the 
inner Nuclear Bulge. With simulations, we estimate 
that there could be at most 40 RRab stars in the MWNSC. 
From the observed RRL star population per unit mass 
in globular clusters, we conclude that an old, metal 
poor (with -1.5$<$[Fe/H]$<$-1) population, 
possibly contributed to the MWNSC by globular 
cluster infall, cannot make up more than about 18\% of  its mass. 
The non-detection of RRab stars from OoII clusters puts an 
even stronger constraint on the fraction of stars 
with even lower (-2.5$<$[Fe/H]$<$-1.5) metallicities: They 
can contribute at most 13\% of the total mass of the MWNSC. 
Of course, old metal poor stars may also have formed in the MWNSC {\sl in situ}. 
This would reduce the contribution to this population from 
globular cluster infall even further. 

%\item We find that only one of our four well-defined 
%RRab stars may actually belong to the MWNSC or lie 
%in the inner Nuclear Bulge. It may have formed {\sl in situ}  
%or have been brought to its current location by an infalling 
%globular cluster. 
%%one of our four well-defined RRab 
%%stars is likely being in the MWNSC, but whether it came from 
%%infall globular star clusters, the inner extent of the Galactic Bulge 
%%or formed {\sl in situ} anciently needs future spectroscopic observations. 
%We have performed simulations that suggest that, in total, there
%  could be at most 40 RRab stars in the MWNSC, after
%  correcting for incompleteness and extinction. From the observed RRL star 
%  population per mass, we
%  conclude 
%that the old, metal poor population, 
%possibly brought to the GC by globular cluster infall with -1.5$<$[Fe/H]$<$-1, 
%does not make up more than 
  %that the globular cluster infall may have contributed not more
  %than 
%  about 18\% of the stars to the MWNSC.

%\item \textbf{The non-detection of RRab star from the OoII clusters puts  
%a strong constrain on the contribution of globular cluster with 
%-2.5$<$[Fe/H]$<$-1.5 on the MWNSC: though the simulation, they 
%can contribute at most 13\% of the total mass of the MWNSC.}

\end{itemize}

\section*{Acknowledgments}
We thank the anonymous referee for a thorough, detailed, and
constructive commentary on our manuscript. 
The research leading to these results has received funding from the
European Research Council under the European Union's Seventh Framework
Programme (FP7/2007-2013) / ERC grant agreement n° [614922]. 
The work is also supported partly 
by NASA
via the grant GO-14589, provided by the Space Telescope Science
Institute. F N-L 
acknowledges financial support from a MECD predoctoral contract, code 
FPU14/01700.
This work
uses observations made with the NASA/ESA Hubble Space Telescope and
the data archive at the Space Telescope Science Institute, which is
operated by the Association of Universities for Research in Astronomy,
Inc. under NASA contract NAS 5-26555.  
We are grateful to Zhiyuan Li, Francisco Najarro, Jon Mauerhan, Farhad Yusef-Zadeh and 
Stephen Eikenberry for
many valuable comments and discussion.

\appendix
\section{Transformation from the Johnson-Cousins-Glass to \hst\ WFC3/IR Magnitudes}
\citet{cat04} give the $M_J$ and $M_H$ of RR Lyrae stars in the Johnson-Cousins-Glass system. In order to 
derive the absolute extinctions and the distance moduli for the RRL stars 
in \S~\ref{ss:extin}, we 
need to translate $M_J$ and $M_H$ into the \hst\ WFC3/IR $M_{F127M}$ and $M_{F153M}$ 
magnitudes. 

We used the stellar atmosphere models by~\citet{cas04}, the transmission curves 
of $J$ and $H$ bands of the Johnson-Cousins-Glass 
system and the F127M and F153M bands of \hst\ WFC3/IR 
distributed in SYNPHOT package provided by STScI to derive the relationship. 
According to~\citet{mar15}, the surface temperature of RR Lyrae stars is 
between 5500 K to 8000 K. Therefore, we only used the stellar atmosphere models 
with this temperature range and also metallicity $\leq0.02$ (solar metallicity) and surface gravity, 
log g $\leq$3. The SYNPHOT package is 
used to derive the intensities of these stellar atmosphere models in units of $Jy$ 
at these four bands. Then, we derived the zero-points in units of $Jy$ from 
the Vega spectrum also provided in the SYNPHOT 
package, which are used to convert the intensities of the stellar atmosphere models 
into magnitudes. We used the least chi-square method to derive the following translation: 
\begin{eqnarray}
M_{F127M}=M_J-0.0059-0.1678\times(M_J-M_H)\\
M_{F153M}=M_H+0.0208+0.0793\times(M_J-M_H)\\
\end{eqnarray}
The difference between $M_{F127M}$ ($M_{F153M}$) and the ones derived from 
the equations above from $M_J$ and $M_H$ are less than 0.008 (0.014) 
mag for the entire range of stellar parameters of RRL stars.
%The median and standard deviation between $M_{F127M}$ ($M_{F153M}$) and the ones
%derived from the equations above from $M_J$ and $M_H$ are -5.8$\times10^{-5}$ (1.6$\times10^{-3}$) 
%and 2.5$\times10^{-3}$ (6.2$\times10^{-3}$) mag. 

\section{Converting $\alpha$ to the relative extinction}
We used the same method given in Appendix A to derive the 
relative extinction $A_{F127M}/A_{Ks}$ and $A_{F153M}/A_{Ks}$ for different $\alpha$. 
We assumed that the temperature and the logarithm surface gravity of RRL stars 
are 6500 $K$ and 2.5 and used the corresponding atmosphere models from 
\citet{cas04}. We first red the spectrum by the extinction curves with different slopes and 
absolute extinctions. Then the `SYNPHOT' package is used to derive the magnitudes at 
the F127M and F153M magnitude, as well as VLT/NACO Ks band. After that, by 
subtracting the magnitudes at these three bands without foreground extinction, respectively, 
we derive the $A_{F127M}$, $A_{F153M}$ and $A_{Ks}$. Finally, we can calculate the mean 
of  $A_{F127M}/A_{Ks}$ and $A_{F153M}/A_{Ks}$ for various foreground extinctions with 
different $\alpha$.

%\begin{figure*}[!thb]
 % \centerline{    %   \epsfig{figure=fig/surface_dis_image.ps,width=0.4\textwidth,angle=0}
 %      \epsfig{figure=fig/dis_ext_com_schultheis.ps,width=1\textwidth,angle=0}
 %      }
 %\caption{The distances derived in \S\ref{ss:extin} vs the $A_K$ derived 
%in \S\ref{ss:extin}. The diamonds represent potential eclipsing binaries 
%due to their large distances. The curve is from the three-dimension extinction map 
%given in~\citet{sch14}. The vertical line marks the location of 
%the MWNSC~\citep{boe16}.}
%\label{f:dis_ext_com}
% \end{figure*}

 %   \begin{figure*}[!thb]
%  \centerline{    %   \epsfig{figure=fig/surface_dis_image.ps,width=0.4\textwidth,angle=0}
%       \epsfig{figure=fig/var_type_paper2.eps,width=1.2\textwidth,angle=0}
%       }
%        \caption{Fourier parameters vs. the logarithm of period. The large symbols are our 21 variable 
%        stars: five-point stars (RRcs), filled triangles (RRabs), open triangles (RRab 
 %       candidates), filled downward triangles (eclipsing binaries) and open filled 
 %       triangles (candidates of eclipsing binaries). The small symbols are RRc  (orange triangles) and 
 %       RRabs (green boxes) from the OGLE survey~\citep{sos09}. 
% }
%\label{f:var_type}
% \end{figure*}

%\begin{figure*}[!thb]
 % \centerline{    %   \epsfig{figure=fig/surface_dis_image.ps,width=0.4\textwidth,angle=0}
  %     \epsfig{figure=fig/rrlyrae_surface.ps,width=1.0\textwidth,angle=0}
   %    }
 %\caption{The surface density of the 1019 RRab stars found in~\citet{gra16}. 
% The solid, dotted and dashed lines are the fitted S\`ersic law with 
% n=1, 2 and 4. The diamond represents the value (3 foreground/background RRab stars) found in our dataset.}
%\label{f:rrlyrae_surface}
% \end{figure*}

\begin{deluxetable}{ccccccccccc}
%\rotate
  \tabletypesize{\tiny}
 \tablecolumns{11}
  \tablecaption{Source Catalog}
   \tablewidth{0pt}
  \tablehead{
     \colhead{} &
     \colhead{} &
    \colhead{} &
    \colhead{} &
    \colhead{} &
    \colhead{} &
    \colhead{} &
    \colhead{} &
    \colhead{$\langle F127M\rangle$} &
    \colhead{} &    
    \colhead{} \\
    \colhead{Name} & 
    \colhead{ID$^a$} &
    \colhead{RA} &
    \colhead{Dec} &
    \colhead{F127M$^{b,c}$} &
    %\colhead{$\sigma_{F127M}$} &
    \colhead{$\langle F127M\rangle$} &
    \colhead{F153M$^{c,d}$} &
    %\colhead{$\sigma_{F153M}^c$} &
    \colhead{$\langle F153M\rangle$} &
    \colhead{-$\langle F153M\rangle^e$} &
    \colhead{Period$^f$}  & 
    \colhead{Type}
}
\startdata
R1&   2495&266.41349&-29.00661&17.2$^{+0.01}_{-0.01}$&17.2&16.6$^{+0.07}_{-0.07}$&16.6&0.56$^{+0.01}_{-0.03}$&0.542&Ecl?\\
R2&   7831&266.42698&-28.99449&20.6$^{+0.04}_{-0.03}$&20.6&17.7$^{+0.10}_{-0.09}$&18.1&2.51$^{+0.02}_{-0.02}$&0.299&RRc\\
R3&   8735&266.42807&-29.00793&20.3$^{+0.07}_{-0.06}$&20.2&18.0$^{+0.13}_{-0.11}$&18.1&2.09$^{+0.04}_{-0.05}$&0.559&RRab\\
R4&   9074&266.42224&-28.99591&20.2$^{+0.02}_{-0.02}$&20.0&18.1$^{+0.11}_{-0.09}$&18.1&1.85$^{+0.02}_{-0.02}$&0.838&Ecl?\\
R5&  10520&266.43683&-29.00501&20.3$^{+0.06}_{-0.05}$&20.4&18.2$^{+0.12}_{-0.11}$&18.3&2.12$^{+0.06}_{-0.07}$&0.542&RRab\\
R6&  12097&266.42163&-28.99880&21.3$^{+0.04}_{-0.03}$&21.3&18.4$^{+0.12}_{-0.10}$&18.5&2.78$^{+0.03}_{-0.03}$&0.489&RRab?/Ecl?\\
R7&  13950&266.39578&-29.00507&21.2$^{+0.11}_{-0.10}$&21.2&18.6$^{+0.16}_{-0.14}$&18.8&2.39$^{+0.06}_{-0.07}$&0.686&RRab?\\
R8&  14123&266.43548&-29.01042&21.3$^{+0.06}_{-0.05}$&21.3&18.6$^{+0.15}_{-0.13}$&18.8&2.53$^{+0.05}_{-0.05}$&0.384&RRc\\
R9&  14583&266.43086&-28.99710&21.3$^{+0.10}_{-0.09}$&21.3&18.8$^{+0.15}_{-0.12}$&18.9&2.37$^{+0.06}_{-0.06}$&0.677&RRab?\\
R10&  15242&266.41267&-29.01958&20.7$^{+0.08}_{-0.07}$&20.7&18.8$^{+0.17}_{-0.14}$&19.0&1.74$^{+0.04}_{-0.04}$&0.329&Ecl?\\
R11&  19440&266.41636&-28.99466&21.5$^{+0.12}_{-0.11}$&21.5&18.9$^{+0.19}_{-0.15}$&19.4&2.02$^{+0.07}_{-0.08}$&0.294&RRc\\
R12&  20993&266.40610&-28.99370&21.7$^{+0.05}_{-0.04}$&21.8&18.9$^{+0.17}_{-0.14}$&19.7&2.17$^{+0.06}_{-0.06}$&0.368&Ecl?\\
R13&  21740&266.41443&-28.99120&22.0$^{+0.09}_{-0.08}$&22.0&19.5$^{+0.21}_{-0.17}$&19.6&2.40$^{+0.07}_{-0.09}$&0.244&RRc\\
R14&  22081&266.42041&-28.99793&22.4$^{+0.08}_{-0.07}$&22.3&19.6$^{+0.17}_{-0.14}$&19.7&2.62$^{+0.07}_{-0.08}$&0.495&RRab?/Ecl?\\
R15&  22197&266.41891&-28.99733&22.8$^{+0.08}_{-0.07}$&22.7&19.7$^{+0.16}_{-0.14}$&19.8&2.96$^{+0.07}_{-0.08}$&0.593&RRab\\
R16&  22312&266.40715&-29.02182&22.4$^{+0.15}_{-0.13}$&22.5&19.8$^{+0.26}_{-0.20}$&19.8&2.63$^{+0.08}_{-0.09}$&0.573&RRab\\
R17&  23037&266.40810&-28.99731&22.2$^{+0.21}_{-0.18}$&22.1&19.8$^{+0.30}_{-0.22}$&20.0&2.13$^{+0.11}_{-0.12}$&0.469&Ecl?\\
R18&  25542&266.42178&-29.01424&22.2$^{+0.24}_{-0.20}$&22.3&19.8$^{+0.30}_{-0.22}$&20.4&1.93$^{+0.14}_{-0.16}$&0.258&Ecl?\\
R19&  25912&266.42983&-29.00620&23.7$^{+0.19}_{-0.16}$&23.7&19.9$^{+0.31}_{-0.22}$&20.5&3.27$^{+0.09}_{-0.11}$&0.644&RRab?\\
R20&  26822&266.42686&-29.01960&22.9$^{+0.11}_{-0.10}$&22.8&20.0$^{+0.19}_{-0.16}$&20.5&2.28$^{+0.08}_{-0.09}$&0.722&Ecl?\\
R21&  28701&266.40033&-28.99777&22.7$^{+0.23}_{-0.19}$&23.0&20.0$^{+0.29}_{-0.21}$&20.9&2.10$^{+0.11}_{-0.13}$&0.235&Ecl?\\
\enddata
\tablecomments{$^a$ ID from the Table 3 of Paper I; $^b$ the F127M magnitudes could 
be different from Table 3 of Paper I, because Table 3 of Paper I gives the mean F127M magnitudes 
of Program GO-11671 and GO-12182, while 
in this table, we primary used the magnitudes from Program 11671, if not, then 
GO-12182; $^c$ The photometric uncertainties are from Table 3 of Paper I, which includes 
both systematic and statistic errors. The former is the bias introduced by crowding, which 
makes the stars artificially brighter; $^d$ from Table 3 of Paper I, the average magnitude 
of the observations 
from 2010 to 2014; $^e$ Because the stellar number density in the F153M band is higher 
than that in the F127M band due to the extinction, the bias introduced by the crowding makes the detected color 
systematically redder; $^f$ in units of days}
\label{t:period}
\end{deluxetable}

\begin{deluxetable}{cccccccccccccc}
  \tabletypesize{\scriptsize}
 \rotate
 \tablecolumns{14}
  \tablecaption{Results of Direct Fourier Fitting}
   \tablewidth{0pt}
  \tablehead{
    \colhead{Name} &
    \colhead{Amplitude$^a$} &
    \colhead{$A_{1}$}&
    \colhead{$\phi_{1}$}&
    \colhead{$A_{21}$}&
    \colhead{$\phi_{21}$}&
    \colhead{$A_{31}$}&
    \colhead{$\phi_{31}$}&
    \colhead{$A_{41}$}&
    \colhead{$\phi_{41}$}&
    \colhead{$A_{51}$}&
    \colhead{$\phi_{51}$}&
    \colhead{$A_{61}$}&
    \colhead{$\phi_{61}$}
}
\startdata
R1& 0.05&0.024&  4.869&0.000&  0.000&0.000&  0.000&0.000&  0.000&0.000&  0.000&0.000&  0.000\\
R2& 0.07&0.049&  3.970&0.353&  4.786&0.000&  0.000&0.000&  0.000&0.000&  0.000&0.000&  0.000\\
R3& 0.10&0.112&  0.411&0.337&  4.169&0.194&  1.823&0.102&  5.680&0.029&  2.481&0.000&  0.000\\
R4& 0.26&0.141&  2.689&0.304&  4.691&0.114&  3.472&0.056&  2.300&0.063&  0.615&0.000&  0.000\\
R5& 0.33&0.128&  4.852&0.392&  3.962&0.250&  1.505&0.189&  5.271&0.101&  2.539&0.097&  6.196\\
R6& 0.33&0.054&  2.373&0.205&  4.569&0.000&  0.000&0.000&  0.000&0.000&  0.000&0.000&  0.000\\
R7& 0.11&0.063&  3.714&0.153&  4.057&0.062&  2.468&0.000&  0.000&0.000&  0.000&0.000&  0.000\\
R8& 0.12&0.049&  1.847&0.000&  0.000&0.000&  0.000&0.000&  0.000&0.000&  0.000&0.000&  0.000\\
R9& 0.13&0.047&  5.282&0.245&  4.270&0.000&  0.000&0.000&  0.000&0.000&  0.000&0.000&  0.000\\
R10& 0.10&0.073&  3.807&0.082&  4.283&0.049&  4.648&0.038&  0.060&0.000&  0.000&0.000&  0.000\\
R11& 0.10&0.086&  4.356&0.168&  4.650&0.000&  0.000&0.000&  0.000&0.000&  0.000&0.000&  0.000\\
R12& 0.15&0.100&  0.498&0.184&  4.554&0.037&  3.352&0.000&  0.000&0.000&  0.000&0.000&  0.000\\
R13& 0.17&0.079&  4.928&0.208&  5.023&0.000&  0.000&0.000&  0.000&0.000&  0.000&0.000&  0.000\\
R14& 0.22&0.081&  4.399&0.145&  5.316&0.000&  0.000&0.000&  0.000&0.000&  0.000&0.000&  0.000\\
R15& 0.21&0.095&  0.802&0.369&  4.367&0.169&  2.029&0.096&  6.114&0.000&  0.000&0.000&  0.000\\
R16& 0.16&0.128&  4.139&0.447&  4.054&0.282&  1.552&0.202&  5.255&0.129&  3.061&0.088&  0.306\\
R17& 0.16&0.132&  1.811&0.206&  4.778&0.090&  2.787&0.041&  1.039&0.000&  0.000&0.000&  0.000\\
R18& 0.22&0.161&  5.156&0.185&  4.799&0.060&  3.970&0.000&  0.000&0.000&  0.000&0.000&  0.000\\
R19& 0.34&0.071&  3.387&0.339&  3.450&0.277&  0.767&0.133&  3.370&0.000&  0.000&0.000&  0.000\\
R20& 0.29&0.067&  0.244&0.417&  5.058&0.329&  3.074&0.375&  1.481&0.372&  0.379&0.275&  5.803\\
R21& 0.25&0.229&  2.877&0.252&  4.800&0.101&  3.215&0.092&  1.348&0.046&  6.030&0.000&  0.000\\
\enddata
\tablecomments{$^a$ in units of mag}
\label{t:rrly}
\end{deluxetable}

\begin{deluxetable}{ccccccccccc}
  \tabletypesize{\tiny}

 \tablecolumns{11}
  \tablecaption{Properties of probable RR Lyrae stars}
   \tablewidth{0pt}
  \tablehead{
     \colhead{}&
     \colhead{}&
     \colhead{}&
     \colhead{$M_{F127M}$-}&
     \colhead{$A_{Ks}^a$}&
   \multicolumn{2}{c}{Slope=-2.2}&
    \multicolumn{2}{c}{Slope=-2.1}&
     \multicolumn{2}{c}{Slope=-2.0}\\ \cmidrule(lr){6-7} \cmidrule(lr){8-9} \cmidrule(lr){10-11}
     \colhead{Name} & 
    \colhead{$M_{F127M}$} &
    \colhead{$M_{F153M}$} &
    \colhead{$M_{F153M}$}&
    \colhead{Red Clump} &
    \colhead{$A_{Ks}^a$} &
    \colhead{Distance$^b$} &
    \colhead{$A_{Ks}^a$} &
    \colhead{Distance$^b$} &
    \colhead{$A_{Ks}^a$} &
    \colhead{Distance$^b$} 
}
\startdata
R1&-0.24$\pm$0.05&-0.39$\pm$0.04& 0.15$\pm$0.01&3.27&0.40$^{+0.10}_{-0.14}$&17.0$^{+1.80}_{-1.31}$&0.44$^{+0.11}_{-0.16}$&16.6$^{+1.86}_{-1
.36}$&0.48$^{+0.12}_{-0.17}$&16.3$^{+1.94}_{-1.41}$\\
R2& 0.24$\pm$0.05& 0.19$\pm$0.04& 0.04$\pm$0.01&2.85&2.41$^{+0.10}_{-0.14}$& 3.8$^{+0.44}_{-0.33}$&2.63$^{+0.11}_{-0.15}$& 3.4$^{+0.41}_{-0
.30}$&2.87$^{+0.12}_{-0.17}$& 2.9$^{+0.37}_{-0.28}$\\
R3&-0.27$\pm$0.05&-0.42$\pm$0.04& 0.15$\pm$0.01&3.42&1.89$^{+0.10}_{-0.14}$& 8.4$^{+0.92}_{-0.68}$&2.06$^{+0.11}_{-0.15}$& 7.7$^{+0.88}_{-0
.65}$&2.25$^{+0.12}_{-0.17}$& 6.9$^{+0.83}_{-0.61}$\\
R4&-0.60$\pm$0.05&-0.82$\pm$0.04& 0.23$\pm$0.01&2.77&1.59$^{+0.10}_{-0.14}$&13.6$^{+1.57}_{-1.19}$&1.73$^{+0.11}_{-0.16}$&12.5$^{+1.52}_{-1
.14}$&1.89$^{+0.12}_{-0.17}$&11.5$^{+1.46}_{-1.09}$\\
R5&-0.24$\pm$0.05&-0.39$\pm$0.04& 0.15$\pm$0.01&3.37&1.93$^{+0.10}_{-0.14}$& 8.6$^{+0.97}_{-0.72}$&2.11$^{+0.11}_{-0.15}$& 7.8$^{+0.92}_{-0
.69}$&2.30$^{+0.12}_{-0.17}$& 7.0$^{+0.87}_{-0.64}$\\
R6&-0.16$\pm$0.05&-0.29$\pm$0.04& 0.13$\pm$0.01&3.00&2.59$^{+0.10}_{-0.14}$& 4.8$^{+0.60}_{-0.46}$&2.83$^{+0.11}_{-0.16}$& 4.2$^{+0.55}_{-0
.41}$&3.09$^{+0.12}_{-0.17}$& 3.6$^{+0.49}_{-0.37}$\\
R7&-0.43$\pm$0.05&-0.62$\pm$0.04& 0.19$\pm$0.01&2.73&2.15$^{+0.10}_{-0.15}$& 9.8$^{+1.21}_{-0.90}$&2.34$^{+0.11}_{-0.16}$& 8.8$^{+1.13}_{-0
.84}$&2.56$^{+0.12}_{-0.17}$& 7.8$^{+1.05}_{-0.77}$\\
R8& 0.04$\pm$0.05&-0.05$\pm$0.04& 0.09$\pm$0.01&3.34&2.39$^{+0.10}_{-0.14}$& 5.9$^{+0.74}_{-0.56}$&2.61$^{+0.11}_{-0.15}$& 5.3$^{+0.68}_{-0
.51}$&2.85$^{+0.12}_{-0.17}$& 4.6$^{+0.62}_{-0.46}$\\
R9&-0.42$\pm$0.05&-0.61$\pm$0.04& 0.19$\pm$0.01&2.42&2.13$^{+0.11}_{-0.15}$&10.4$^{+1.40}_{-1.05}$&2.32$^{+0.12}_{-0.16}$& 9.3$^{+1.31}_{-0
.97}$&2.54$^{+0.13}_{-0.18}$& 8.2$^{+1.21}_{-0.90}$\\
R10& 0.16$\pm$0.05& 0.10$\pm$0.04& 0.06$\pm$0.01&3.54&1.65$^{+0.10}_{-0.14}$&12.3$^{+1.56}_{-1.16}$&1.79$^{+0.11}_{-0.16}$&11.3$^{+1.50}_{-
1.11}$&1.96$^{+0.12}_{-0.17}$&10.3$^{+1.42}_{-1.05}$\\
R11& 0.25$\pm$0.05& 0.21$\pm$0.04& 0.04$\pm$0.01&3.35&1.94$^{+0.11}_{-0.15}$&11.0$^{+1.58}_{-1.18}$&2.11$^{+0.12}_{-0.17}$&10.0$^{+1.49}_{-
1.11}$&2.31$^{+0.13}_{-0.18}$& 8.9$^{+1.38}_{-1.03}$\\
R12& 0.07$\pm$0.05&-0.01$\pm$0.04& 0.08$\pm$0.01&3.61&2.05$^{+0.11}_{-0.15}$&12.1$^{+1.54}_{-1.17}$&2.23$^{+0.12}_{-0.16}$&10.9$^{+1.45}_{-
1.09}$&2.44$^{+0.13}_{-0.18}$& 9.7$^{+1.35}_{-1.01}$\\
R13& 0.40$\pm$0.05& 0.40$\pm$0.04& 0.01$\pm$0.01&3.38&2.34$^{+0.11}_{-0.16}$& 7.5$^{+1.21}_{-0.89}$&2.55$^{+0.12}_{-0.17}$& 6.6$^{+1.10}_{-
0.81}$&2.79$^{+0.13}_{-0.18}$& 5.8$^{+1.00}_{-0.73}$\\
R14&-0.17$\pm$0.05&-0.30$\pm$0.04& 0.13$\pm$0.01&2.86&2.43$^{+0.11}_{-0.15}$& 9.9$^{+1.73}_{-1.26}$&2.65$^{+0.12}_{-0.16}$& 8.8$^{+1.57}_{-
1.13}$&2.89$^{+0.13}_{-0.18}$& 7.6$^{+1.40}_{-1.01}$\\
R15&-0.32$\pm$0.05&-0.48$\pm$0.04& 0.16$\pm$0.01&3.33&2.73$^{+0.12}_{-0.17}$& 8.2$^{+1.46}_{-1.05}$&2.98$^{+0.13}_{-0.18}$& 7.1$^{+1.30}_{-
0.94}$&3.25$^{+0.14}_{-0.20}$& 6.1$^{+1.15}_{-0.83}$\\
R16&-0.29$\pm$0.05&-0.45$\pm$0.04& 0.16$\pm$0.01&2.92&2.42$^{+0.11}_{-0.16}$&11.2$^{+2.04}_{-1.45}$&2.64$^{+0.12}_{-0.17}$& 9.9$^{+1.85}_{-
1.31}$&2.88$^{+0.13}_{-0.19}$& 8.7$^{+1.66}_{-1.17}$\\
R17&-0.13$\pm$0.05&-0.25$\pm$0.04& 0.12$\pm$0.01&3.24&1.97$^{+0.11}_{-0.15}$&16.7$^{+2.30}_{-1.77}$&2.14$^{+0.12}_{-0.17}$&15.1$^{+2.17}_{-
1.66}$&2.34$^{+0.13}_{-0.18}$&13.5$^{+2.02}_{-1.54}$\\
R18& 0.36$\pm$0.05& 0.34$\pm$0.04& 0.02$\pm$0.01&3.13&1.87$^{+0.11}_{-0.15}$&16.8$^{+3.04}_{-2.21}$&2.04$^{+0.12}_{-0.17}$&15.3$^{+2.83}_{-
2.05}$&2.22$^{+0.13}_{-0.18}$&13.8$^{+2.61}_{-1.89}$\\
R19&-0.38$\pm$0.05&-0.56$\pm$0.04& 0.18$\pm$0.01&3.22&3.03$^{+0.10}_{-0.14}$& 8.8$^{+1.67}_{-1.20}$&3.30$^{+0.11}_{-0.16}$& 7.6$^{+1.46}_{-
1.05}$&3.61$^{+0.12}_{-0.17}$& 6.4$^{+1.25}_{-0.90}$\\
R20&-0.48$\pm$0.05&-0.67$\pm$0.04& 0.20$\pm$0.01&3.29&2.04$^{+0.10}_{-0.15}$&24.5$^{+3.72}_{-2.92}$&2.22$^{+0.11}_{-0.16}$&22.1$^{+3.46}_{-
2.70}$&2.43$^{+0.12}_{-0.18}$&19.7$^{+3.18}_{-2.47}$\\
R21& 0.43$\pm$0.05& 0.43$\pm$0.04& 0.00$\pm$0.01&3.15&2.05$^{+0.13}_{-0.18}$&17.1$^{+3.28}_{-2.51}$&2.24$^{+0.14}_{-0.19}$&15.4$^{+3.03}_{-
2.31}$&2.44$^{+0.15}_{-0.21}$&13.7$^{+2.78}_{-2.11}$\\

\enddata
\tablecomments{$^a$, in units of mag; $^b$, in units of kpc}
\label{t:rrly_pro}
\end{deluxetable}

\end{document}